\pdfoutput=1
\documentclass[pdflatex,sn-mathphys-num]{sn-jnl}

\usepackage{graphicx}%
\usepackage{multirow}%
\usepackage{amsmath,amssymb,amsfonts}%
\usepackage{amsthm}%
\usepackage{mathrsfs}%
\usepackage[title]{appendix}%
\usepackage{xcolor}%
\usepackage{textcomp}%
\usepackage{manyfoot}%
\usepackage{booktabs}
\usepackage{booktabs,threeparttable}%
\usepackage{adjustbox}%
\usepackage{algorithm}%
\usepackage{algorithmicx}%
\usepackage{algpseudocode}%
\usepackage{listings}%
\usepackage{soul} 
\usepackage{cancel}
\usepackage{rotating} 
\usepackage{url}
\usepackage{booktabs}   
\usepackage{adjustbox}  
\usepackage{amsmath}    
\DeclareGraphicsExtensions{.pdf,.png,.jpg}

\usepackage{amsmath,amssymb,bm}   
\usepackage{booktabs}             
\usepackage{array}                
\usepackage{threeparttable}       
\usepackage{adjustbox}            
\usepackage{graphicx}             
\usepackage{textcomp}             
\usepackage{siunitx}              
\usepackage{siunitx}       
\usepackage{booktabs}       
\usepackage{amsmath}        
\usepackage{threeparttable} 
\usepackage{siunitx}        
\usepackage{placeins}       
\usepackage{booktabs}
\usepackage{threeparttable}
\usepackage{siunitx}
\theoremstyle{thmstyleone}%
\theoremstyle{thmstyletwo}%

\theoremstyle{thmstylethree}%

\begin{document}

\title[Bayesian Semiparametric Joint Modeling of Gap-Time Distribution for Multitype Recurrent Events and a Terminal Event]{Bayesian Semiparametric Joint Modeling of Gap-Time Distribution for Multitype Recurrent Events and a Terminal Event}


\author*[1]{\fnm{Mithun Kumar} \sur{Acharjee}}\email{acharjee@uab.edu}

\author[1]{\fnm{AKM Fazlur} \sur{Rahman}}

\affil[1]{\orgdiv{Department of Biostatistics}, \orgname{University of Alabama at Birmingham}, \orgaddress{\street{1665 University Blvd}, \city{Birmingham}, \postcode{35294}, \state{Alabama}, \country{USA}}}


\abstract{In biomedical settings, multitype recurrent events such as stroke and heart failure occur frequently, often concluding with a terminal event such as death. Understanding the links between these recurring and terminal events is fundamental to developing interventions that delay detrimental outcomes. Joint modeling is needed to quantify the dependence between event types and between recurrent events and mortality. We proposed a Bayesian semiparametric joint model on the gap–time scale for \emph{multitype} recurrent events and a terminal event. The model includes a \emph{shared frailty} that links all recurrent types and the terminal event. Each baseline hazard is assigned a \emph{gamma–process} prior, while regression and frailty parameters receive standard parametric priors. This ensures flexible baselines and familiar effect measures. The construction gives closed–form expressions for the cumulative hazards function and frailty component and connects to Breslow-Aalen type estimators as a special case of our proposed estimator. This links our Bayesian procedure to the classical approach. Computationally, we developed a simple MCMC sampler that avoids large matrix factorizations and scales nearly linearly in the sample size. The comprehensive simulation evaluates four criteria: \emph{accuracy, prediction, robustness, and computation}. There is no exact frequentist version of our specification; for comparison, we fit the same model with an EM algorithm in a frequentist framework. Our model and MCMC algorithm demonstrated superior performance in each criterion. We illustrate the approach with data from the antihypertensive and lipid–lowering treatment to prevent heart attack trial (ALLHAT), jointly analyzing acute and chronic cardiovascular recurrences and death.}




\maketitle
\section{Introduction}\label{sec:intro}
Recurrent events are repeated outcomes experienced by the same individual and arise widely in biomedical research. Examples include repeated thrombosis and infections in dialysis patients \cite{ravani2013temporal}, shunt failures in pediatric hydrocephalus \cite{tuli2000risk}, and tumor recurrences in bladder cancer \cite{byar1980veterans}. For a single recurrence type, established approaches include complete-intensity counting-process models \cite{andersen1982cox,andersen2012statistical}, marginal rate models \cite{pepe1993some,lawless1995some,lin2000semiparametric}, and gap-time formulations that model the waiting time between events \cite{pena2001nonparametric,pena2007semiparametric,rahman2014nonparametric}. Joint models that link a single recurrent process with a terminal event are also well developed \cite{zhangsheng2011joint,yu2014joint,ghosh2000nonparametric,li2019bayesian}; these use either marginal modeling \cite{cook2007statistical,ghosh2000nonparametric} or frailty-based formulations (single shared frailty or cause-specific frailty) \cite{huang2004joint,huang2007joint,rondeau2007joint}. Compared with type-specific frailties, a shared subject-level frailty captures cross-type dependence with a single latent factor, yielding a more parsimonious and interpretable model \cite{huang2004joint,huang2007joint}. At the same time, many works fit frailty models on the calendar-time scale with piecewise-constant or penalized baselines estimated via EM or penalized likelihood \cite{cook2007statistical,hougaard2000analysis,ripatti2000estimation,duchateau2008frailty,mclachlan2008em}.

In more complex scenarios, subjects may experience multiple types of recurrent events along with a terminal event. Notable studies, such as the Danish Heart Failure Study \cite{kober1995clinical}, the Framingham Heart Study \cite{kannel1979diabetes}, the Women’s Health Initiative \cite{writing2002risks}, and the Atherosclerosis Risk in Communities (ARIC) Study \cite{aric1989atherosclerosis}, highlight the importance of jointly modeling multitype recurrent events (MTREs) and a terminal event. These event types often interact, with one type influencing the likelihood of others, making independent analyses potentially misleading \cite{ghosh2023dynamic,lin2017bayesian}. For example, if the terminal event is death and the recurrent events include heart attacks and strokes, each type may be correlated with mortality. This situation demands joint models that accurately capture this hidden heterogeneity.

Relatively little research has addressed joint modeling of \emph{multitype} recurrent events and a terminal event. Lin et al.\ \cite{lin2017bayesian} developed a Bayesian model for multitype recurrent events with dependent termination on the calendar-time scale, using spline-based (piecewise-constant) baseline hazards. Liu and Pe\~na \cite{liu2015dynamic} proposed a frequentist dynamic model for multitype recurrent events and a terminal event fitted by the EM algorithm, also on calendar time. In simulations, the latter fails to estimate the frailty parameter robustly (bias $\approx 1.06$), in line with the well-known sensitivity of EM fits to starting values \cite{mclachlan2008em,ibrahim2001}. The piecewise-constant baseline used by \cite{lin2017bayesian} may compromise smoothness and trend recovery \cite{jachno2021impact}, and the Nelson–Aalen–type baseline in \cite{liu2015dynamic} is a step function that can be rough and unstable with sparse events \cite{pena1993small}.

We develop a Bayesian semiparametric joint shared–frailty model on the \emph{gap-time} scale, with gamma–process priors for the baseline hazard, and fit it via a simple MCMC algorithm. Working on gap time resets the clock after each event, directly modeling the wait to the next recurrence and cleanly separating within-person dynamics from calendar-time effects \cite{prentice1981regression,cook2007statistical,amorim2015modelling,kelly2000survival}. This helps avoid distorted risk evaluation and supports individualized strategies \cite{wang1999nonparametric,huang2007joint,bailey2015estimation}, while remaining consistent with counting-process foundations \cite{andersen1982cox,prentice1981regression}. The gamma–process prior yields smooth, data-adaptive hazards that improve stability over stepwise estimators \cite{li2021bayesian}, and full posterior inference naturally quantifies uncertainty where EM-based approaches are fragile \cite{ibrahim2001}.

In simulations, we fit the model with an adaptive Metropolis–Hastings MCMC algorithm. The sampler exploits the low-dimensional summaries implied by the gamma–process prior, keeping likelihood evaluations lightweight and avoiding large matrix factorizations or quadrature. Computation splits across subjects, so each iteration scales nearly linearly with sample size. We assess \emph{accuracy}, \emph{prediction}, \emph{robustness}, and \emph{computational efficiency}. Using the same specification as our EM implementation, the Bayesian fit performs well across criteria, whereas the EM fit shows the expected sensitivity to starting values and requires additional work for standard errors \cite{mclachlan2008em,ibrahim2001}.
\begin{figure*}[!ht]\label{fig1}
\centering
\includegraphics[width=\textwidth]{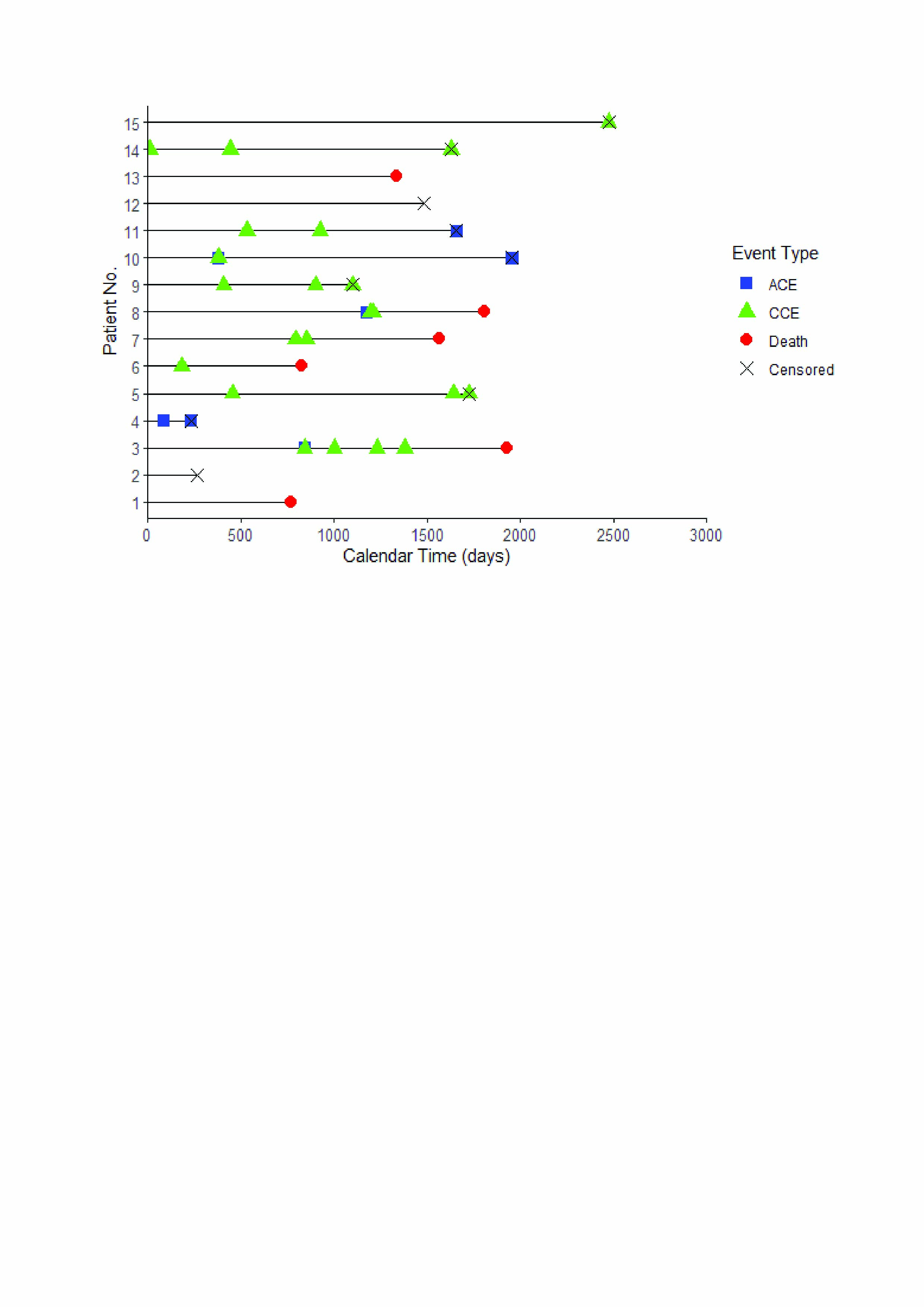}
\vspace{-4in}
\caption{Multitype recurrent events and a terminal event for 15 selected participants from the ALLHAT trial.}
\label{fig:Figure_1_Mithun.jpg}
\end{figure*}
 \FloatBarrier 
We apply our model to the ALLHAT clinical trial dataset \cite{davis1996rationale} (Figure \ref{fig:Figure_1_Mithun.jpg}) to analyze acute cardiovascular events (ACE: myocardial infarction and stroke), chronic cardiovascular events (CCE: congestive heart failure, angina, and peripheral arterial disease), and death, leveraging the model’s ability to capture event dependencies and treatment effects. By including a frailty term, we assess the association between recurrent events and the terminal event. We examine the impact of age, race (Black vs.\ non-Black), and antihypertensive treatments (Amlodipine and Lisinopril vs.\ Chlorthalidone as the reference drug) on ACE, CCE, and death. Hazard ratios directly compare the three treatment groups for each outcome. The model also estimates baseline cumulative hazards and visualizes treatment-specific survival probabilities, yielding practical insights into the nature of MTREs and a terminal event.

The proposed model is mathematically rigorous and ensures accurate inference by using a closed-form cumulative hazard while capturing event dependencies on the gap-time scale. In addition, we provide a closed-form estimator of subject-specific frailties. Compared with frequentist approaches, our framework (i) yields general cumulative hazard estimates that recover Breslow–Aalen–type nonparametric estimators and (ii) combines high estimation accuracy with computational efficiency. The ALLHAT dataset illustrates the model’s practical utility. All analyses were performed in \textsf{R}, using custom functions we developed.

The remainder of this article is organized as follows. Section \ref{sec:method} introduces notation and the model, derives the likelihood and conditional posterior distributions, and describes the MCMC algorithm. Section \ref{sec:sim} presents extensive simulation studies. Section \ref{sec:appli} illustrates the application to the ALLHAT clinical trial. Concluding remarks are provided in Section \ref{sec:diss}.

\section{Methodology}\label{sec:method}

\subsection{Joint Frailty Model}
Consider \(n\) subjects in the study and let \(W_1, W_2, \dots, W_n\) be independent and identically distributed (IID) positive-valued frailties from a parametric distribution \(H(w; \nu)\). More precisely, the \(W_i\) are assumed IID Gamma with unit mean and variance \(1/\nu\); that is, \(W_i \sim \mathrm{Gamma}(\nu, \nu)\), \(i=1,2,\dots,n\). The mean of \(W\) is set to 1 to make the frailty parameter \(\nu\) identifiable.

Given the unobserved frailty variable \(W_i\), we assume that the successive inter-event times (gap-times) for the type-\(q\) (\(q=1,2,\dots,Q\)) recurrent event of subject \(i\), denoted by \(\{T_{qij},\, j=1,2,\dots\}\), are IID nonnegative random variables with a common distribution function
\[
\bar{F}_q(t \mid W = w)
= \big[\bar{F}_{0q}(t)\big]^w
= \exp\{-w\,\Lambda_{0q}(t)\},
\]
where \(\bar{F}_{0q}(t)\) and \(\Lambda_{0q}(t) = \int_0^t \lambda_{0q}(u)\,du\) with \(\lambda_{0q}(t) = \frac{dF_{0q}(t)}{\bar{F}_{0q}(t)}\) are the baseline survival function (SF) and baseline cumulative hazard function (CHF), respectively. Conditionally on \(W\), the gap-times \(\{T_{qij}\}\) are independent, but unconditionally they are dependent.

The \(i\)th subject is observed over \([0,\tau_i]\), where \(\tau_1,\tau_2,\dots,\tau_n\) are IID with common distribution function \(G\). It is assumed that \(\tau_i\) and \(\{T_{qij}, j=1,2,\dots\}\) are mutually independent. The (marginal) survival function of \(\{T_{qij}\}\) for type-\(q\) recurrent events is therefore
\[
\bar{F}_q(t) = \mathbb{E}\big[\bar{F}_{0q}(t)^W\big]
= \left[\frac{\nu}{\nu + \Lambda_{0q}(t)}\right]^\nu.
\]
Smaller \(\nu\) indicates stronger correlation between gap-times; larger \(\nu\) indicates weaker correlation. For instance, under exponential baseline with \(\Lambda_{01}(t)=\eta t\),
\[
\mathrm{Corr}(T_{111}, T_{112}) = 1/\nu;\ \nu>2.
\]

Let \(\mathbf{X}=(X_1, X_2, \dots, X_p)\) be a \(p\)-dimensional observable covariate vector. In the presence of covariates \(\mathbf{X}=\mathbf{x}\), given \(W=w\), we assume \(\{T_{qij}\}_{j\ge 1}\) are IID with
\[
\bar{F}_q(t \mid W=w, \mathbf{X}=\mathbf{x})
= \big[\bar{F}_{0q}(t)\big]^{\,w\exp(\beta_q^\top\mathbf{X})}
= \exp\!\left[-\,w\,\exp(\beta_q^\top\mathbf{X})\,\Lambda_{0q}(t)\right],
\]
where \(\beta_q\) is the parameter vector for type-\(q\) events.

Let \(T_{0i}\) be the terminal-event time for subject \(i\), where \(T_{01},T_{02},\dots,T_{0n}\) are IID with common distribution \(H_0\). The baseline and conditional survival functions of the terminal event are
\begin{equation*}
\begin{aligned}
\bar{F}_0(t) &= \mathbb{E}\big[\bar{F}_{00}(t)^W\big]
= \left[\frac{\nu}{\nu + \Lambda_{00}(t)}\right]^\nu, \\
\bar{F}_0(t \mid W=w,\mathbf{X}=\mathbf{x})
&= \big[\bar{F}_{00}(t)\big]^{\,w\exp(\beta_0^\top\mathbf{X})}
= \exp\!\left[-\,w\,\exp(\beta_0^\top\mathbf{X})\,\Lambda_{00}(t)\right],
\end{aligned}
\end{equation*}
where \(\beta_0\) is the parameter vector for the terminal event; \(\bar{F}_{00}(t)\) and \(\Lambda_{00}(t)=\int_0^t \lambda_{00}(u)\,du\) with \(\lambda_{00}(t)=\frac{dF_{00}(t)}{\bar{F}_{00}(t)}\) are the baseline SF and baseline CHF.

Therefore, we consider the joint intensity for type-\(q\) recurrent events and the terminal event:
\begin{equation}
\left\{
\begin{aligned}
\lambda_{qi}(t \mid W_i,\mathbf{X}_i) &= W_i\,\lambda_{0q}(t)\,\exp(\beta_q^\top \mathbf{X}_i), \quad q=1,2,\dots,Q, \\
\lambda_{0i}(t \mid W_i,\mathbf{X}_i) &= W_i\,\lambda_{00}(t)\,\exp(\beta_0^\top \mathbf{X}_i),
\end{aligned}
\right.
\label{eq1}
\end{equation}
where \(\lambda_{0q}(t)\) and \(\lambda_{00}(t)\) are the baseline hazards; \(\lambda_{qi}(t)\) and \(\lambda_{0i}(t)\) are the corresponding subject-specific hazards.

For subject \(i\), the numbers of observed events for the type-\(q\) recurrent processes and the terminal event are
\begin{equation*}
\begin{aligned}
K_{qi} &= \max\big\{k \in \{0,1,\ldots\}: S_{qik} \le \min(\tau_i, T_{0i})\big\}, \\
K_{0i} &= I(T_{0i} \le \tau_i),
\end{aligned}
\end{equation*}
where \(S_{qi0}=0\), \(S_{qik}=\sum_{j=1}^k T_{qij}\) for \(k=1,2,\dots\). The observable random vector is
\[
D_{qi}^* = \big(\tau_i, T_{0i}, K_{qi}, K_{0i}, T_{qi1}, T_{qi2}, \ldots, T_{qiK_{qi}}, \min(\tau_i, T_{0i}) - S_{qiK_{qi}}\big).
\]

\subsection{Likelihood}
Let \( B_{qi}(u) = u - S_{qiN_{qi}^\dagger(u^-)} \) be the backward recurrence time for the type-\(q\) events. In the stochastic process method, $Y_{qi}^\dagger (s)$ (\( s \geq 0 \)) is the \textit{at-risk process} for type-\(q\) recurrent events indicating subject \(i\) is still under observation for type-\(q\) events at time \(s\), $N_{qi}^\dagger (s)$ is the \textit{counting process} which counts the number of failures due to cause \(q\) for $i^{th}$ subject during calendar time period [0, \(s\)], $Y_{0i}^\dagger (s)$ (\( s \geq 0 \)) is the \textit{at-risk process} for a terminal event indicating subject \(i\) is still under observation for a terminal event at time \(s\), $N_{0i}^\dagger (s)$ is the \textit{counting process} which is an indicator function denote whether we observed the event  (1) or not (0). We can define them as follows:
\begin{equation*}
\begin{aligned}
    Y_{qi}^\dagger (s) &= I \{\tau_i \ge \min(s, T_{0i})\} \nonumber \\
    N_{qi}^\dagger (s) &= \sum_{j = 1}^{\infty} I\{S_{qij} \leq s, S_{qij} \leq \min(\tau_i, T_{0i})\} \nonumber \\
    Y_{0i}^\dagger (s) &= I \{\min(\tau_i, T_{0i}) \ge s\} \nonumber \\
    N_{0i}^\dagger (s) &=  I\{T_{0i} \leq \min(s, \tau_i)\} \nonumber
\end{aligned}
\end{equation*}

Let \( D_i^* = \{D_{qi}^*\}_{q=1}^Q \) and \( D^* = \{D_{i}^*, \mathbf{X}_i\}_{i=1}^n \). Then, following Jacod \cite{jacod1975multivariate} or Section II.7 of Andersen et al. \cite{andersen2012statistical}, and using Eq.~(\ref{eq1}), the conditional joint full likelihood over \([0,s]\) is
\begin{align}
L(s \mid D^*, w_1,\dots,w_n)
&= \prod_{i=1}^{n} \prod_{q=1}^{Q}
\Bigg[
\Bigg(\prod_{u \in [0,s]}
\big[W_i \exp(\beta_q^\top \mathbf{X}_i)\, Y_{qi}^\dagger(u)\, \lambda_{0q}(B_{qi}(u))\big]^{N_{qi}^\dagger(\Delta u)}\Bigg)
\nonumber\\[-2pt]
&\qquad\qquad\times
\exp\!\left\{-W_i \exp(\beta_q^\top \mathbf{X}_i) \int_0^s Y_{qi}^\dagger(u)\, \lambda_{0q}(B_{qi}(u))\,du\right\}
\Bigg]
\nonumber\\
&\quad\times
\prod_{i=1}^{n}
\Bigg[
\Bigg(\prod_{u \in [0,s]}
\big[W_i \exp(\beta_0^\top \mathbf{X}_i)\, Y_{qi}^\dagger(u)\, \lambda_{00}(u)\big]^{N_{0i}^\dagger(\Delta u)}\Bigg)
\nonumber\\[-2pt]
&\qquad\qquad\times
\exp\!\left\{-W_i \exp(\beta_0^\top \mathbf{X}_i) \int_0^s Y_{qi}^\dagger(u)\, \lambda_{00}(u)\,du\right\}
\Bigg].
\label{eq2}
\end{align}

Since in our Bayesian procedure we assign gamma-process priors to \(\Lambda_{0q}(\cdot)\) and \(\Lambda_{00}(\cdot)\), it is convenient to simplify the likelihood for posterior computation. Following Peña et al. \cite{pena2000weak},
\begin{equation*}
\begin{aligned}
\int_0^s Y_{qi}^\dagger(u)\, \lambda_{0q}(B_{qi}(u))\,du
&=  \int_{0}^{s} Y_{i}(s,u)\, \lambda_{0q}(u)\,du,
\end{aligned}
\end{equation*}
where
\begin{equation}
Y_{i}(s,u)
= \sum_{j=1}^{N_{qi}^\dagger((s\cap \tau_i)^-)} I(T_{qij} \ge u)
+ I\big((s\cap \tau_i)- S_{qiN_{qi}^\dagger((s\cap \tau_i)^-)} \ge u\big)\nonumber.
\end{equation}
When \(s \to \infty\),
\begin{equation}
Y_{i}(s,u) \to \sum_{j=1}^{K_{qi}} I(T_{qij} \ge u) + I(\tau_i - S_{qiK_{qi}} \ge u) \equiv Y_{i}(u),\nonumber
\end{equation}
which holds true when \(s \ge \tau_i\).

Let \(t_{(1)}, t_{(2)}, \dots, t_{(M)}\) partition \(\mathbb{R}_+ = (0,\infty)\) with \(t_{(0)}\equiv0\) and \(t_{(M+1)}\equiv\infty\) so that \(Y_{i}(s,u)\) is constant within each sub-interval \((t_{(j-1)}, t_{(j)}]\). Define
\[
\Lambda_{0q}(\Delta t_{(j)}) = \Lambda_{0q}(t_{(j)}) - \Lambda_{0q}(t_{(j-1)}),
\quad j=1,2,\dots,M,M+1;\ q=1,2,\dots,Q.
\]
Then
\[
\int_{0}^{s} Y_{i}(s,u)\,\lambda_{0q}(u)\,du
= \sum_{j=1}^{M+1} Y_{i}(s,t_{(j)})\, \Lambda_{0q}(\Delta t_{(j)}).
\]
Similarly, the product integral can be written as
\[
\prod_{u \in [0,s]} \big[Y_{qi}^\dagger(u)\,\lambda_{0q}(B_{qi}(u))\big]^{N_{qi}^\dagger(\Delta u)}
= \prod_{j=1}^{M+1} \big[Y_{qi}^\dagger(t_{(j)})\, \Lambda_{0q}(\Delta t_{(j)})\big]^{N_{qi}^\dagger(\Delta t_{(j)})},
\]
where \(N_{qi}^\dagger(\Delta t_{(j)}) = N_{qi}^\dagger(t_{(j)}) - N_{qi}^\dagger(t_{(j-1)})\).

Let \(s \ge \max\{\tau_1,\dots,\tau_n\}\). Using the above notation, the joint likelihood in Eq.~(\ref{eq2}) becomes
\begin{align}
& L\big(\Lambda_{0q}(\cdot), \Lambda_{00}(\cdot), \beta_q, \beta_0, \nu \mid D^*, w_1,\dots,w_n\big)
\nonumber\\
&\quad=
\prod_{i=1}^{n} \prod_{q=1}^{Q}
\Bigg[
\Bigg(\prod_{j=1}^{M+1}
\big[W_i \exp(\beta_q^\top \mathbf{X}_i)\, Y_{qi}^\dagger(t_{(j)})\, \Lambda_{0q}(\Delta t_{(j)})\big]^{N_{qi}^\dagger(\Delta t_{(j)})}\Bigg)
\nonumber\\[-2pt]
&\qquad\qquad\times
\exp\!\left\{-W_i \exp(\beta_q^\top \mathbf{X}_i)\, \sum_{j=1}^{M+1} Y_{qi}^\dagger(t_{(j)})\, \Lambda_{0q}(\Delta t_{(j)})\right\}
\Bigg]
\nonumber\\
&\qquad\times
\prod_{i=1}^{n}
\Bigg[
\Bigg(\prod_{j=1}^{M+1}
\big[W_i \exp(\beta_0^\top \mathbf{X}_i)\, Y_{qi}^\dagger(t_{(j)})\, \Lambda_{00}(\Delta t_{(j)})\big]^{N_{0i}^\dagger(\Delta t_{(j)})}\Bigg)
\nonumber\\[-2pt]
&\qquad\qquad\times
\exp\!\left\{-W_i \exp(\beta_0^\top \mathbf{X}_i)\, \sum_{j=1}^{M+1} Y_{qi}^\dagger(t_{(j)})\, \Lambda_{00}(\Delta t_{(j)})\right\}
\Bigg].
\label{eq3}
\end{align}

\subsection{Prior Specifications and Conditional Posteriors}
The unknown parameters of interest are \(\Lambda_{0q}(\cdot), \Lambda_{00}(\cdot), \nu, \beta_q, \beta_0\); however, we also derive the conditional distribution of the unobservable frailty vector \(W_1, W_2, \dots, W_n\), as those will be involved in the Bayes estimator of \(\Lambda_{0q}(\cdot)\) and \(\Lambda_{00}(\cdot)\). Following Kalbfleisch \cite{kalbfleisch1978non}, we assume \(\Lambda_{0q}(\cdot)\) and \(\Lambda_{00}(\cdot)\) have a nonparametric gamma-process prior \(\Lambda_{0q}(\cdot) \sim \mathcal{G}_{c_q,\Lambda_{0q}^*(\cdot)}\)
and \(\Lambda_{00}(\cdot) \sim \mathcal{G}_{c_0,\Lambda_{00}^*(\cdot)}\)
where \(\Lambda_{0q}^*(\cdot)\) and \(\Lambda_{00}^*(\cdot)\) are completely known mean intensity functions for type-\(q\) recurrent events and a terminal event, respectively. Also, \(c_q\) and \(c_0\) represent the precision of the prior measure for recurrent and terminal events, respectively. Then, we can write
\[
\Lambda_{0q}(\Delta t_{(j)}) \sim \mathrm{Gamma}\!\big(c_q\,\Lambda_{0q}^*(\Delta t_{(j)}),\, c_q\big)
\quad\text{and}\quad
\Lambda_{00}(\Delta t_{(j)}) \sim \mathrm{Gamma}\!\big(c_0\,\Lambda_{00}^*(\Delta t_{(j)}),\, c_0\big).
\]
We use the notation \(\pi (\Lambda_{0q}(\Delta t_{(j)}))\) to denote the prior distribution of \(\Lambda_{0q}(\Delta t_{(j)})\) and so on. That is,
\begin{equation*}
\begin{aligned}
    \pi(\Lambda_{0q}(\Delta t_{(j)})) & \sim  \mathrm{Gamma}\!\big(c_q\Lambda_{0q}^*(\Delta t_{(j)}),\,c_q\big), \quad j = 1, 2, \dots, M + 1, \\
    \pi(\Lambda_{00}(\Delta t_{(j)})) & \sim  \mathrm{Gamma}\!\big(c_0\Lambda_{00}^*(\Delta t_{(j)}),\,c_0\big), \quad j = 1, 2, \dots, M + 1.
\end{aligned}
\end{equation*}
We assume the prior of \(\nu\) as a Gamma distribution with a known shape \(\zeta\) and rate \(\eta\). Also, the prior for \(\beta_q\) and \(\beta_0\) are specified as \(p\)-dimensional multivariate normal distributions with known mean vectors \(\mu_{\beta_q}\) and \(\mu_{\beta_0}\), and corresponding variance–covariance matrices \(\Sigma_{\beta_q}\) and \(\Sigma_{\beta_0}\), respectively. We define the prior distributions
\begin{equation*}
\begin{aligned}
    \pi(\nu) & \sim \mathrm{Gamma}(\zeta, \eta),\\
    \pi(\beta_q) & \sim N_p(\mu_{\beta_q}, \Sigma_{\beta_q}),\\
    \pi(\beta_0) & \sim N_p(\mu_{\beta_0}, \Sigma_{\beta_0}).
\end{aligned}
\end{equation*}
Using Eq.~(\ref{eq3}) and the above-specified priors, we define the joint posterior distribution of \(\{\Lambda_{0q}(\cdot), \Lambda_{00}(\cdot), \nu, \beta_q, \beta_0\}\) via
\begin{align}
    p(\Lambda_{0q}(\cdot),\Lambda_{00}(\cdot), W, \nu, \beta_q, \beta_0 \mid D^*) &\propto L(\Lambda_{0q}(\cdot), \Lambda_{00}(\cdot), \beta_q, \beta_0, \nu \mid D^*, W_1, W_2,\dots, W_n)\nonumber\\
    &\mathrel{\times} \pi(\Lambda_{0q}(\cdot))\,\pi(\Lambda_{00}(\cdot))\,\pi(\nu)\,\pi(\beta_q)\,\pi(\beta_0).
\label{eq4}
\end{align}

Let
\begin{equation*}
\begin{aligned}
    r_{qi} \equiv   \exp(\beta_q^\top \mathbf{X}_i)\sum_{j=1}^{M+1} \Big[Y_{qi}^\dagger (t_{(j)}) \Lambda_{0q} (\Delta t_{(j)})\Big],\\
    r_{0i} \equiv  \exp(\beta_0^\top \mathbf{X}_i)\sum_{j=1}^{M+1} \Big[Y_{qi}^\dagger (t_{(j)}) \Lambda_{00} (\Delta t_{(j)})\Big].
\end{aligned}
\end{equation*}

Then, the conditional posterior distributions of \(\Lambda_{0q}(\cdot)\) and \(\Lambda_{00}(\cdot)\) are as follows:
\begin{align}
    &p(\Lambda_{0q} (\Delta t_{(j)})\mid\Lambda_{00} (\Delta t_{(j)}), \mathbf{w}, \nu, \beta_q,\beta_0) \propto \mathrm{Gamma}\Bigg(\sum_{i=1}^{n}  N_{qi}^\dagger (\Delta t_{(j)}) + c_q \cdot \Lambda_{0q}^* (\Delta t_{(j)}), \notag\\
    &\hspace{4em} c_q + \sum_{i=1}^{n}  W_i \exp(\beta_q^\top \mathbf{X}_i) Y_{qi}^\dagger (t_{(j)})\Bigg),
\label{eq5}\\
    &p(\Lambda_{00} (\Delta t_{(j)})\mid\Lambda_{0q} (\Delta t_{(j)}), \mathbf{w}, \nu, \beta_q,\beta_0) \propto \mathrm{Gamma}\Bigg(\sum_{i=1}^{n}  N_{0i}^\dagger (\Delta t_{(j)}) + c_0 \cdot \Lambda_{00}^* (\Delta t_{(j)}), \notag\\
    &\hspace{4em} c_0 + \sum_{i=1}^{n}  W_i \exp(\beta_0^\top \mathbf{X}_i) Y_{qi}^\dagger (t_{(j)})\Bigg).
\label{eq6}
\end{align}
Under an integrated squared-error loss function, we obtain
\begin{align}
    \widetilde{\Lambda}_{0q}(t\mid\mathbf{W},\Lambda_{00}, \nu, \beta_q,\beta_0) &= \sum_{j=1}^{M+1} \left[ \frac{\sum_{i=1}^{n} N_{qi}^\dagger (\Delta t_{(j)}) + c_q \cdot \Lambda_{0q}^* (\Delta t_{(j)})}{c_q + \sum_{i=1}^{n} W_i \exp(\beta_q^\top \mathbf{X}_i) Y_{qi}^\dagger (t_{(j)})} \right] I(t_{(j)} \leq t),
\label{eq7}\\
    \widetilde{\Lambda}_{00}(t\mid\mathbf{W},\Lambda_{0q}, \nu, \beta_q,\beta_0) &= \sum_{j=1}^{M+1} \left[ \frac{\sum_{i=1}^{n} N_{0i}^\dagger (\Delta t_{(j)}) + c_0 \cdot \Lambda_{00}^* (\Delta t_{(j)})}{c_0 + \sum_{i=1}^{n} W_i \exp(\beta_0^\top \mathbf{X}_i) Y_{qi}^\dagger (t_{(j)})} \right] I(t_{(j)} \leq t).
\label{eq8}
\end{align}
Here \(\mathbf{W}=(W_1,\dots,W_n)\). These are not yet estimators of \(\Lambda_{0q}(t)\) and \(\Lambda_{00}(t)\) since the \(W_i\) are unknown. However, we can obtain \(\hat{W}_i\) to replace \(W_i\) using the following conditional posterior distribution.
\begin{align}
    p(W_i \mid \Lambda_{0q} (.), \Lambda_{00} (.), \nu, \beta_q, \beta_0) &\propto W_i^{N_{. i}^\dagger (.) + N_{0i}^\dagger (.)} \exp(-W_i (r_{.i} + r_{0i})) \times g_{W_i}(w_i \mid \nu) \notag \\
    &\quad \propto \mathrm{Gamma} \Big(N_{.i}^\dagger (.) + N_{0i}^\dagger (.) + \nu, r_{.i} + r_{0i} + \nu \Big),
\label{eq9}
\end{align}
where \(N_{. i}^\dagger (.) = \sum_{q=1}^{Q} N_{qi}^\dagger (.)\), \(r_{.i} = \sum_{q=1}^{Q} r_{qi}\), and \( g_{W_i}(w_i\mid\nu)  \propto W_i^{\nu - 1} \exp\!\left(-W_i \nu\right)\). The posterior mean is
 \begin{equation}
     \hat{W}_i = \frac{N_{.i}^\dagger (.) + N_{0i}^\dagger (.) +\nu}{r_{.i} + r_{0i} + \nu}, \quad i=1,2,\dots,n.
     \label{eq10}
 \end{equation}
Thus, we have a closed-form estimator of \(\Lambda_{0q}(t)\) and \(\Lambda_{00}(t)\) given by
\begin{align}
    \hat{\Lambda}_{0q}(t\mid\mathbf{W},\Lambda_{00}, \nu, \beta_q,\beta_0) &= \sum_{j=1}^{M+1} \left[ \frac{\sum_{i=1}^{n} N_{qi}^\dagger (\Delta t_{(j)}) + c_q \cdot \Lambda_{0q}^* (\Delta t_{(j)})}{c_q + \sum_{i=1}^{n} \hat{W}_i \exp(\beta_q^\top \mathbf{X}_i) Y_{qi}^\dagger (t_{(j)})} \right] I(t_{(j)} \leq t),
\label{eq11}\\
    \hat{\Lambda}_{00}(t\mid\mathbf{W},\Lambda_{0q}, \nu, \beta_q,\beta_0) &= \sum_{j=1}^{M+1} \left[ \frac{\sum_{i=1}^{n} N_{0i}^\dagger (\Delta t_{(j)}) + c_0 \cdot \Lambda_{00}^* (\Delta t_{(j)})}{c_0 + \sum_{i=1}^{n} \hat{W}_i \exp(\beta_0^\top \mathbf{X}_i) Y_{qi}^\dagger (t_{(j)})} \right] I(t_{(j)} \leq t).
\label{eq12}
\end{align}
We can recover the Breslow–Aalen–type estimator of the baseline cumulative hazard function from Eqs.~(\ref{eq11}) and (\ref{eq12}) by letting the precision parameters \( c_q \to 0 \) and \( c_0 \to 0 \). The conditional posterior distribution of the frailty parameter \( \nu \) is given by
\begin{align}
    p(\nu \mid \Lambda_{0q} (.), \Lambda_{00} (.), \beta_0, \beta_q) &\propto L_m (\Lambda_{0q} (.), \Lambda_{00} (.), \nu, \beta_q, \beta_0) \times \pi(\nu) \notag \\
    &\propto \prod_{i=1}^{n} \Bigg[ \frac{\Gamma (N_{.i}^\dagger (.) + N_{0i}^\dagger (.) + \nu)}{(r_{.i} + r_{0i} + \nu)^{(N_{.i}^\dagger (.) + N_{0i}^\dagger (.) + \nu)}} \times \frac{\nu^\nu}{\Gamma(\nu)} \Bigg] \times \pi(\nu),
\label{eq13}
\end{align}
where \(L_m\) denotes the marginal likelihood and \(\pi(\nu) \propto \nu^{\zeta-1} \exp(-\eta \nu)\). The conditional posterior distributions of the regression parameters \(\beta_q\) and \(\beta_0\) are given by
\begin{align}
    p(\beta_q \mid \Lambda_{0q} (.), \Lambda_{00} (.), \mathbf{W}, \nu, \beta_0) &\propto L_m (\beta_q, \Lambda_{0q} (.), \Lambda_{00} (.), \nu, \beta_0) \times \pi(\beta_q) \notag \\
    &\propto \exp \Bigg[ \sum_{i=1}^{n} N_{qi}^\dagger (.) \cdot (\beta_q^\top \mathbf{X}_i) - \sum_{i=1}^{n} W_i \cdot r_{qi} \Bigg] \times \pi(\beta_q),
\label{eq14}
\end{align}
\begin{align}
    p(\beta_0 \mid \Lambda_{0q} (.), \Lambda_{00} (.), \mathbf{W}, \nu,  \beta_q) &\propto L_m (\beta_0, \Lambda_{0q} (.), \Lambda_{00} (.), \nu, \beta_q) \times \pi(\beta_0) \notag \\
    &\propto \exp \Bigg[ \sum_{i=1}^{n} N_{0i}^\dagger (.) \cdot (\beta_0^\top \mathbf{X}_i) - \sum_{i=1}^{n} W_i \cdot r_{0i} \Bigg] \times \pi(\beta_0),
\label{eq15}
\end{align}
where \(\pi(\beta_q) \propto \exp\Big[-\frac{1}{2} (\beta_q-\mu_{\beta_q})^\top\Sigma_{\beta_q}^{-1} (\beta_q-\mu_{\beta_q})\Big]\) and \(\pi(\beta_0) \propto \exp\Big[-\frac{1}{2} (\beta_0-\mu_{\beta_0})^\top\Sigma_{\beta_0}^{-1} (\beta_0-\mu_{\beta_0})\Big]\).

The inter-event-time baseline survival functions for type-\(q\) recurrent events and the terminal event can be estimated via
\begin{align}
\hat{\bar{F}}_q(t) &= \left[\frac{\hat{\nu}}{\hat{\nu} + \hat{\Lambda}_{0q}(t)}\right]^{\hat{\nu}},
\label{eq16}
\end{align}
\begin{align}
\hat{\bar{F}}_0(t) &= \left[\frac{\hat{\nu}}{\hat{\nu} + \hat{\Lambda}_{00}(t)}\right]^{\hat{\nu}},
\label{eq17} 
\end{align}
where \(\hat{\Lambda}_{0q}(t)\) and \(\hat{\Lambda}_{00}(t)\) are defined in Eqs.~(\ref{eq11}) and (\ref{eq12}), respectively. Also, \(\hat{\nu}\) is an estimator of \(\nu\) obtained using Eq.~(\ref{eq13}) as outlined in the \emph{MCMC Algorithm} subsection.

\subsection{MCMC Algorithm}
The conditional posteriors of \(\Lambda_{0q}\), \(\Lambda_{00}\), and \(W_i\) in Eqs.~(\ref{eq5}), (\ref{eq6}), and (\ref{eq9}) are available in closed form and can be sampled directly or updated via the Bayes estimators in Eqs.~(\ref{eq7}), (\ref{eq8}), and (\ref{eq10}). For \(\nu\), \(\beta_q\), and \(\beta_0\), we employ Metropolis–Hastings (MH) updates targeting Eqs.~(\ref{eq13}), (\ref{eq14}), and (\ref{eq15}), respectively.

\paragraph{Pseudocode}
\begin{enumerate}
\item \textbf{Initialization:} Set \(\Lambda_{0q}^{(0)}(\cdot), \Lambda_{00}^{(0)}(\cdot), W^{(0)}, \nu^{(0)}, \beta_q^{(0)}, \beta_0^{(0)}\).
\item \textbf{For} \(m=1,\dots,M\) \textbf{do}
  \begin{enumerate}
    \item For each \(q\), update \(\Lambda_{0q}^{(m)}\) either by drawing increments using Eq.~(\ref{eq5}) or by computing \(\widetilde{\Lambda}_{0q}\) via Eq.~(\ref{eq7}) with current \(W^{(m-1)}\) and \(\beta_q^{(m-1)}\).
    \item Update \(\Lambda_{00}^{(m)}\) analogously using Eq.~(\ref{eq6}) or Eq.~(\ref{eq8}) with current \(W^{(m-1)}\) and \(\beta_0^{(m-1)}\).
    \item Compute \(r_{qi}\) and \(r_{0i}\) using the current \(\Lambda_{0q}^{(m)}\), \(\Lambda_{00}^{(m)}\), and \(\beta_q^{(m-1)}, \beta_0^{(m-1)}\) (definitions preceding Eq.~(\ref{eq5})).
    \item \emph{Sample} \(\nu^{(m)}\) by Metropolis–Hastings targeting the conditional posterior in Eq.~(\ref{eq13})).
    \item Update \(W_i^{(m)}\) for \(i=1,\dots,n\) either by drawing from Eq.~(\ref{eq9}) or by the posterior mean in Eq.~(\ref{eq10}).
    \item For each \(q\), \emph{sample} \(\beta_q^{(m)}\) from the conditional posterior given in Eq.~(\ref{eq14}) (via Metropolis–Hastings).
    \item \emph{Sample} \(\beta_0^{(m)}\) from the conditional posterior given in Eq.~(\ref{eq15}) (via Metropolis–Hastings).
  \end{enumerate}
\item After burn-in and thinning, compute \(\hat{\Lambda}_{0q}\) and \(\hat{\Lambda}_{00}\) via Eqs.~(\ref{eq11})–(\ref{eq12}), and \(\hat{\bar{F}}_q, \hat{\bar{F}}_0\); summarize posterior draws for point estimates, credible intervals, and diagnostics.
\end{enumerate}

\section{Simulation Study}\label{sec:sim}

We conducted Monte Carlo experiments to evaluate the finite-sample performance of our Bayesian semiparametric joint frailty model on the gap-time scale and to benchmark it against a frequentist EM-based comparator. We first specify the design in full (mechanism, priors, censoring, initialization, scenarios, and comparator), then report results focusing on accuracy, prediction, robustness, and computation.

\subsection{Simulation Design}

\subsubsection{Data-generation mechanism}

\paragraph{Model and hazards.}
For each subject \(i=1,\dots,n\), we first draw a shared frailty \(W_i\sim\mathrm{Gamma}(\nu,\nu)\) with \(\nu\in\{2,4\}\) so that \(\mathbb{E}(W_i)=1\) and \(\mathrm{Var}(W_i)=1/\nu\). We then generate two independent covariates, \(X_{i1}\sim\mathrm{Bernoulli}(0.5)\) and \(X_{i2}\sim\mathcal{N}(0,1)\), and set \(X_i=(X_{i1},X_{i2})^\top\). The process includes two recurrent types (\(q=1,2\)) and one terminal event (\(q=0\)). Conditional gap-time hazards follow the multiplicative structure in Eq.~(\ref{eq1}), with shared frailty \(W_i\), baseline hazards \(\lambda_{0q}(t)\) for the two recurrent types and \(\lambda_{00}(t)\) for the terminal event, and type-specific regression vectors \(\beta_1=(-0.4,\,0.35)^\top\), \(\beta_2=(-0.3,\,0.25)^\top\), and \(\beta_0=(-0.1,\,0.1)^\top\). Baseline hazards are Weibull with a common shape \(\gamma\) and type-specific scales \((\lambda_1,\lambda_2,\lambda_0)=(1.2,\,1.1,\,3.2)\); we study both increasing-failure-rate (IFR; \(\gamma=1.1\)) and decreasing-failure-rate (DFR; \(\gamma=0.9\)) regimes.

\paragraph{Event times and censoring.}
At each gap origin, for components \(q\in\{1,2,0\}\) (terminal event \(q=0\)), we draw independent candidate gaps using a Weibull baseline with shape \(\gamma\) and scale \(\lambda_q\in\{\lambda_1,\lambda_2,\lambda_0\}\) and multiplicative factor \(M_{iq}=W_i\exp(\beta_q^\top X_i)\): \(T_q=\lambda_q\{(-\log U_q)/M_{iq}\}^{1/k}\) with \(U_q\sim\mathrm{Unif}(0,1)\). Here \(k\) denotes the Weibull shape parameter (equal to \(\gamma\) in our notation). Independent censoring times are \(\tau_i\sim\mathrm{Unif}(1,3)\) with administrative end at \(\tau=3\). The next event time is \(\min(T_1,T_2,T_0)\); we realize the corresponding type, reset the gap-time clock, and repeat until whichever occurs first: terminal event or censoring time. 

\paragraph{Bayesian priors.}
Each cumulative baseline hazard \(\Lambda_{0q}(t)\) is modeled with independent Gamma–process increments on a working grid \(\{t_{(j)}\}\):
\[
\Lambda_{0q}(\Delta t_{(j)})\;\sim\;\mathrm{Gamma}\!\Big(c_q\,\Lambda_{0q}^\ast(\Delta t_{(j)}),\;c_q\Big),
\]
with reference means
\[
\Lambda_{01}^\ast(t)=\Big(\tfrac{t}{1.1}\Big)^\gamma,\quad
\Lambda_{02}^\ast(t)=\Big(\tfrac{t}{1.0}\Big)^\gamma,\quad
\Lambda_{00}^\ast(t)=\Big(\tfrac{t}{3.1}\Big)^\gamma,
\]
and baseline precision \(c_q=c_0=0.1\) (weakly informative). Regression priors are \(\beta_{q\ell}\sim\mathcal{N}(0,1)\) where \(\ell = 1,2\), and the frailty prior is \(\nu\sim\mathrm{Gamma}(2,2)\). For robustness, we additionally (i) conduct a \emph{baseline-prior misspecification} analysis by replacing the Weibull-based reference means with exponential working means, i.e., \(\Lambda_{01}^\ast(t)=t/\lambda_1\), \(\Lambda_{02}^\ast(t)=t/\lambda_2\), and \(\Lambda_{00}^\ast(t)=t/\lambda_0\), paired with a lower precision \(c_q=c_0=0.01\), and summarize results via RMSE plots of estimated inter-event-time baseline survival functions (Eqs.~(\ref{eq16})–(\ref{eq17})); and (ii) perform a \emph{frailty-parameter prior sensitivity} check by re-fitting with a log-normal prior on the frailty parameter \(\nu\) (in place of the \(\mathrm{Gamma}(\zeta,\eta)\) prior), moment-matched to preserve the Gamma prior’s mean and variance: with \(\mu_g=\zeta/\eta\) and \(\mathrm{Var}_g=\zeta/\eta^2\), set \(s_\nu^2=\log\!\big(1+\mathrm{Var}_g/\mu_g^2\big)\) and \(m_\nu=\log(\mu_g)-s_\nu^2/2\), i.e., \(\nu\sim\mathrm{LogNormal}(m_\nu,s_\nu^2)\). The conditional frailty remains \(W_i\mid \nu \sim \mathrm{Gamma}(\nu,\nu)\), with the same regression priors and baseline-process specification.

\paragraph{Initialization.}
Each regression vector has two coefficients and the frailty parameter is a single scalar. We initialize the type-specific regression vectors at \(\beta_1^{(0)}=(-1.5,\,1.5)^\top\), \(\beta_2^{(0)}=(-1.5,\,1.5)^\top\), \(\beta_0^{(0)}=(-1.8,\,1.9)^\top\), and set the frailty parameter to \(\nu^{(0)}=3\). Baseline cumulative-hazard increments on the partition points \(\{t_{(j)}\}=\{0,0.03,\dots,3\}\) start at their prior means, i.e., \(\Lambda_{0q}^{(0)}(\Delta t_{(j)})=\Lambda_{0q}^\ast(\Delta t_{(j)})\) for \(q\in\{1,2,0\}\).
For robustness, we use two \emph{fixed} alternative starts while keeping the baselines at their prior means in both cases: (i) an ``all-ones'' start with \(\beta_1^{(0)}=(1,1)^\top\), \(\beta_2^{(0)}=(1,1)^\top\), \(\beta_0^{(0)}=(1,1)^\top\), \(\nu^{(0)}=1\); and (ii) an ``all-fours'' start with \(\beta_1^{(0)}=(4,4)^\top\), \(\beta_2^{(0)}=(4,4)^\top\), \(\beta_0^{(0)}=(4,4)^\top\), \(\nu^{(0)}=4\).

\paragraph{Scenarios and replication.}
We vary (i) sample size \(n\in\{100,300\}\), (ii) frailty parameter \(\nu\in\{2,4\}\) in the generator, and (iii) Weibull shape \(\gamma\in\{1.1,0.9\}\) (IFR/DFR). For each scenario, we simulate \(500\) datasets. Bayesian fits use \(5{,}000\) iterations, burn-in \(2{,}000\), and a thinning of \(5\). Identical random seeds are used across replications and across methods (Bayesian vs.\ frequentist) to ensure exact reproducibility and fair comparisons. 

\paragraph{Frequentist comparator.}
There is \emph{no exact frequentist analogue} of our Bayesian semiparametric prior on \(\Lambda_{0q}\); as a benchmark we adopt the closest widely used multiplicative-frailty proportional hazard formulation and estimate its \emph{cumulative hazard functions} on the gap-time scale via Breslow-type sums. With at-risk indicators \(Y_{qi}^\dagger(t)\), counting increments \(N_{qi}^\dagger(\Delta t_{(j)})\) over \(\Delta t_{(j)}\!=\!(t_{(j-1)},t_{(j)}]\), and E–step weights \(\hat W_i\), the recurrent-type (\(q=1,2\)) estimator is
\begin{equation}
\label{eq:freq-ch-rec-compact}
\hat{\Lambda}_{0q}^{\boldsymbol f}(t)
=\sum_{j}\!\left[
\frac{\sum_{i=1}^{n} N_{qi}^\dagger(\Delta t_{(j)})}
{\sum_{i=1}^{n} \hat W_i\,\exp(\beta_q^\top X_i)\,Y_{qi}^\dagger(t_{(j)})}
\right]\mathbb{I}\{t_{(j)}\le t\},
\end{equation}
and the terminal-event estimator is
\begin{equation}
\label{eq:freq-ch-term-compact}
\hat{\Lambda}_{00}^{\boldsymbol f}(t)
=\sum_{j}\!\left[
\frac{\sum_{i=1}^{n} N_{0i}^\dagger(\Delta t_{(j)})}
{\sum_{i=1}^{n} \hat W_i\,\exp(\beta_0^\top X_i)\,Y_{0i}^\dagger(t_{(j)})}
\right]\mathbb{I}\{t_{(j)}\le t\}.
\end{equation}
Here the superscript \({\boldsymbol f}\) denotes the \emph{frequentist} cumulative-hazard estimators (for recurrent types in \eqref{eq:freq-ch-rec-compact} and for the terminal event in \eqref{eq:freq-ch-term-compact}). All frequentist fits use \emph{exactly} the same simulated datasets, random seeds, partition points, covariates, and censoring scheme as the Bayesian analysis to ensure a reproducible comparison.

\subsubsection{Model Evaluation Criteria}

\paragraph{Accuracy.}
Across replications, we summarize estimation performance using four metrics: (i) bias, the difference between the average estimate and the truth; (ii) standard deviation, the empirical variability of the estimates; (iii) root mean squared error, which combines squared bias and variance; and (iv) empirical coverage of nominal 95\% intervals for both Bayesian credible intervals and frequentist confidence intervals (Tables~\ref{tab:gamma1.1-noEst}–\ref{tab:gamma0.9-noEst}).

\paragraph{Prediction.}
\begin{enumerate}
\item \emph{Prediction via RMSE (Figure~\ref{fig:Figure_2_Mithun.pdf}).} Using Eqs.~(\ref{eq16})–(\ref{eq17}), we computed the pointwise RMSE of estimated baseline survival curves \(\hat{\bar{F}}_q(t)\) and \(\hat{\bar{F}}_0(t)\) to compare Bayesian vs.\ frequentist prediction.
\item \emph{Survival overlays (Figure~\ref{fig:Figure_5_Mithun.pdf}).} We compute the estimated survival curve for a type-\(q\) event \(\hat S_q(t)=\exp\{-\hat{\Lambda}_{0q}(t)\}\) (and the terminal counterpart) and overlay it with the true survival \(S_q(t)\) to assess how closely the model recovers the true survival function.
\end{enumerate}

\paragraph{Robustness.}
\begin{enumerate}
\item \emph{Convergence diagnostics (Table~\ref{tab:diagnostics}).} For a single replicated dataset, we run \(4\) chains with \(10{,}000\) iterations per chain, \(5{,}000\) burn-in, and thinning \(=5\); we report Gelman–Rubin \(\widehat R\) and bulk/tail effective sample sizes (ESS) based on \(4{,}000\) drawn samples to assess parameter convergence.
\item \emph{Baseline–prior misspecification (Figure~\ref{fig:Figure_3_Mithun.pdf}).} Replace Weibull-based reference means \(\Lambda_{0q}^\ast(t)=(t/\lambda_q)^\gamma\) with exponential working means \(\Lambda_{0q}^\ast(t)=t/\lambda_q\) using \((\lambda_1,\lambda_2,\lambda_0)=(6.2,\,7.1,\,8.2)\), and set \(c_q=c_0=0.01\); report RMSE of \(\hat{\bar{F}}_q(t)\) and \(\hat{\bar{F}}_0(t)\) to evaluate the impact of misspecification.
\item \emph{Informativeness sensitivity (Figure~\ref{fig:Figure_4_Mithun.pdf}).} We vary prior informativeness for the regressions and for the frailty parameter \(\nu\) across four tiers. For \(\nu\sim\mathrm{Gamma}(\zeta,\eta)\) with \(\zeta=\eta\), we have \(\mathbb{E}(\nu)=1\) and \(\sigma_\nu^2=\mathrm{Var}(\nu)=\zeta/\eta^{2}=1/\eta\). The tiers are: strong \((\sigma_\beta^{2}=0.5,\ \zeta=\eta=4 \Rightarrow \sigma_\nu^{2}=0.25)\), standard \((\sigma_\beta^{2}=1,\ \zeta=\eta=2 \Rightarrow \sigma_\nu^{2}=0.5)\), weak \((\sigma_\beta^{2}=2.25,\ \zeta=\eta=1 \Rightarrow \sigma_\nu^{2}=1)\), and vague \((\sigma_\beta^{2}=9,\ \zeta=\eta=0.5 \Rightarrow \sigma_\nu^{2}=2)\). Pointwise RMSE is computed using Eqs.~(\ref{eq16})–(\ref{eq17}) to assess sensitivity to prior informativeness. 
\item \emph{Prior sensitivity for the frailty parameter (Table~\ref{tab:prior_sensitivity_four}).} Re-fit with a moment-matched log-normal prior on \(\nu\) and compare against the Gamma-prior fit. 
\item \emph{Initialization sensitivity (Table~\ref{tab:Rk_two_scenarios}).} Use two fixed, commonly chosen alternative starts (e.g., ``all-ones'' and ``all-fours'' for each two-dimensional \(\beta_q\) with \(\nu\in\{1,4\}\)), keeping the baseline increments at prior means; compare resulting parameter estimates to assess robustness and to compare Bayesian behavior vs.\ frequentist behavior under different initializations. 
\end{enumerate}

\paragraph{Computation.}

\emph{Computation time (Table~\ref{tab:runtime}).} Summary statistics for each replicated dataset and scenario are reported to assess the computational gain of the Bayesian model over the frequentist comparator.

\subsection{Simulation Results}

Across all scenarios, the Bayesian estimator recovers the regression effects with smaller RMSE than the EM fit and with comparable or better coverage. For the regression effects \((\beta_{11},\beta_{12},\beta_{21},\beta_{22})\), the Bayesian RMSE advantage persists: for example, at \(\gamma=1.1,n=100\) the Bayesian RMSEs are \(0.168,0.113,0.160,0.117\) versus EM \(0.253,0.326,0.274,0.299\) (Table~\ref{tab:gamma1.1-noEst}). Estimating the frailty parameter \(\nu\) is intrinsically difficult because the likelihood surface is often flat and the parameter is weakly identified in moderate samples \cite{hougaard1995,ripatti2000,therneau2003}. As a result, both methods perform worse for \(\nu\) than for the regression effects. Despite this challenge, the Bayesian fit is consistently more accurate for \(\nu\) than EM across all scenarios. For example, at \(\gamma=1.1,n=100,\nu=2\) the Bayesian RMSE/CP are \(0.423/0.99\) versus EM \(1.614/0.20\) (Table~\ref{tab:gamma1.1-noEst}); at \(\gamma=0.9,n=100,\nu=4\) the Bayesian RMSE is \(0.549\) versus EM \(3.565\) (Table~\ref{tab:gamma0.9-noEst}); and at \(n=300,\nu=2\) the Bayesian RMSE is \(0.474\text{--}0.910\) versus EM \(1.597\text{--}1.613\) across \(\gamma\in\{0.9,1.1\}\) (Tables~\ref{tab:gamma1.1-noEst}–\ref{tab:gamma0.9-noEst}).

\begin{table*}[!t]
\centering
\begin{threeparttable}
\caption{Simulation results comparing Bayesian and frequentist model based on increasing failure rate ($\gamma = 1.1$).}
\label{tab:gamma1.1-noEst}
\renewcommand{\arraystretch}{1.3}
\setlength{\tabcolsep}{4pt}
\begin{tabular*}{\textwidth}{@{\extracolsep\fill}ccccccccccc}
\toprule
\multicolumn{3}{@{}c@{}}{\textbf{Scenario}} &
\multicolumn{4}{@{}c@{}}{\textbf{Bayesian Model: MCMC}} &
\multicolumn{4}{@{}c@{}}{\textbf{Frequentist Model: EM}} \\
\cmidrule(lr){1-3}\cmidrule(lr){4-7}\cmidrule(lr){8-11}
\textbf{n} & $\boldsymbol{\nu}$ & \textbf{Parameter} & \textbf{Bias} & \textbf{SD} & \textbf{RMSE} & \textbf{CP} & \textbf{Bias} & \textbf{SD} & \textbf{RMSE} & \textbf{CP} \\
\midrule
100 & 2 & $\nu$       & -0.223 & 0.360 & 0.423 & 0.99 & 1.614 & 0.047 & 1.614 & 0.20 \\
100 & 2 & $\beta_{11}$ & -0.079 & 0.148 & 0.168 & 0.90 & -0.138 & 0.212 & 0.253 & 0.75 \\
100 & 2 & $\beta_{12}$ &  0.006 & 0.113 & 0.113 & 0.89 & -0.226 & 0.235 & 0.326 & 0.70 \\
100 & 2 & $\beta_{21}$ & -0.077 & 0.141 & 0.160 & 0.92 & -0.168 & 0.217 & 0.274 & 0.74 \\
100 & 2 & $\beta_{22}$ &  0.005 & 0.117 & 0.117 & 0.91 & -0.189 & 0.231 & 0.299 & 0.72 \\
100 & 2 & $\beta_{01}$ & -0.074 & 0.192 & 0.206 & 0.96 & -0.409 & 0.332 & 0.527 & 0.60 \\
100 & 2 & $\beta_{02}$ & -0.001 & 0.171 & 0.171 & 0.94 & -0.040 & 0.256 & 0.259 & 0.88 \\
100 & 4 & $\nu$       &  0.919 & 0.413 & 1.008 & 0.80 & 3.574 & 0.063 & 3.575 & 0.15 \\
100 & 4 & $\beta_{11}$ & -0.059 & 0.119 & 0.133 & 0.95 & -0.141 & 0.215 & 0.257 & 0.80 \\
100 & 4 & $\beta_{12}$ & -0.011 & 0.096 & 0.097 & 0.93 & -0.246 & 0.241 & 0.345 & 0.72 \\
100 & 4 & $\beta_{21}$ & -0.060 & 0.121 & 0.136 & 0.96 & -0.176 & 0.219 & 0.281 & 0.80 \\
100 & 4 & $\beta_{22}$ & -0.009 & 0.108 & 0.108 & 0.91 & -0.205 & 0.237 & 0.313 & 0.70 \\
100 & 4 & $\beta_{01}$ & -0.054 & 0.167 & 0.175 & 0.98 & -0.410 & 0.360 & 0.545 & 0.62 \\
100 & 4 & $\beta_{02}$ & -0.007 & 0.166 & 0.166 & 0.93 & -0.039 & 0.261 & 0.264 & 0.88 \\
300 & 2 & $\nu$       & -0.381 & 0.282 & 0.474 & 0.92 & 1.612 & 0.043 & 1.613 & 0.18 \\
300 & 2 & $\beta_{11}$ & -0.105 & 0.089 & 0.137 & 0.78 & -0.280 & 0.224 & 0.359 & 0.58 \\
300 & 2 & $\beta_{12}$ &  0.051 & 0.065 & 0.083 & 0.79 & -0.133 & 0.240 & 0.274 & 0.68 \\
300 & 2 & $\beta_{21}$ & -0.084 & 0.083 & 0.118 & 0.84 & -0.275 & 0.228 & 0.357 & 0.60 \\
300 & 2 & $\beta_{22}$ &  0.042 & 0.069 & 0.081 & 0.86 & -0.127 & 0.235 & 0.268 & 0.74 \\
300 & 2 & $\beta_{01}$ & -0.068 & 0.108 & 0.128 & 0.95 & -0.438 & 0.335 & 0.552 & 0.62 \\
300 & 2 & $\beta_{02}$ &  0.001 & 0.103 & 0.103 & 0.92 & -0.032 & 0.253 & 0.255 & 0.88 \\
300 & 4 & $\nu$       & -0.335 & 0.600 & 0.687 & 0.98 & 3.567 & 0.065 & 3.568 & 0.20 \\
300 & 4 & $\beta_{11}$ & -0.077 & 0.076 & 0.108 & 0.87 & -0.256 & 0.209 & 0.331 & 0.65 \\
300 & 4 & $\beta_{12}$ &  0.037 & 0.058 & 0.069 & 0.85 & -0.147 & 0.249 & 0.289 & 0.72 \\
300 & 4 & $\beta_{21}$ & -0.071 & 0.072 & 0.101 & 0.85 & -0.270 & 0.212 & 0.344 & 0.64 \\
300 & 4 & $\beta_{22}$ &  0.032 & 0.057 & 0.065 & 0.89 & -0.141 & 0.244 & 0.282 & 0.74 \\
300 & 4 & $\beta_{01}$ & -0.038 & 0.099 & 0.106 & 0.98 & -0.427 & 0.310 & 0.527 & 0.60 \\
300 & 4 & $\beta_{02}$ &  0.002 & 0.096 & 0.096 & 0.93 & -0.037 & 0.260 & 0.263 & 0.88 \\
\bottomrule
\end{tabular*}
\begin{tablenotes}[flushleft]\footnotesize
\item Bias: mean(estimate$-$true); SD: empirical standard deviation; RMSE: root mean squared error; CP: nominal 95\% coverage probability.
\end{tablenotes}
\end{threeparttable}
\end{table*}
\FloatBarrier

\begin{table*}[!t]
\centering
\begin{threeparttable}
\caption{Simulation results comparing Bayesian and frequentist model based on decreasing failure rate ($\gamma = 0.9$).}
\label{tab:gamma0.9-noEst}
\renewcommand{\arraystretch}{1.3}
\setlength{\tabcolsep}{4pt}
\begin{tabular*}{\textwidth}{@{\extracolsep\fill}ccccccccccc}
\toprule
\multicolumn{3}{@{}c@{}}{\textbf{Scenario}} &
\multicolumn{4}{@{}c@{}}{\textbf{Bayesian Model: MCMC}} &
\multicolumn{4}{@{}c@{}}{\textbf{Frequentist Model: EM}} \\
\cmidrule(lr){1-3}\cmidrule(lr){4-7}\cmidrule(lr){8-11}
\textbf{n} & $\boldsymbol{\nu}$ & \textbf{Parameter} & \textbf{Bias} & \textbf{SD} & \textbf{RMSE} & \textbf{CP} & \textbf{Bias} & \textbf{SD} & \textbf{RMSE} & \textbf{CP} \\
\midrule
100 & 2 & $\nu$       & -1.016 & 0.402 & 1.093 & 0.80 & 1.595 & 0.056 & 1.596 & 0.15 \\
100 & 2 & $\beta_{11}$ & -0.099 & 0.119 & 0.155 & 0.92 & -0.177 & 0.154 & 0.235 & 0.80 \\
100 & 2 & $\beta_{12}$ &  0.077 & 0.100 & 0.126 & 0.82 & -0.139 & 0.247 & 0.283 & 0.70 \\
100 & 2 & $\beta_{21}$ & -0.054 & 0.125 & 0.136 & 0.98 & -0.125 & 0.157 & 0.201 & 0.90 \\
100 & 2 & $\beta_{22}$ &  0.032 & 0.117 & 0.121 & 0.86 & -0.169 & 0.244 & 0.297 & 0.72 \\
100 & 2 & $\beta_{01}$ & -0.072 & 0.169 & 0.184 & 0.98 & -0.207 & 0.169 & 0.267 & 0.84 \\
100 & 2 & $\beta_{02}$ &  0.119 & 0.170 & 0.207 & 0.80 & -0.035 & 0.233 & 0.236 & 0.76 \\
100 & 4 & $\nu$       &  0.273 & 0.477 & 0.549 & 0.98 & 3.564 & 0.067 & 3.565 & 0.20 \\
100 & 4 & $\beta_{11}$ & -0.071 & 0.128 & 0.147 & 0.84 & -0.185 & 0.183 & 0.260 & 0.70 \\
100 & 4 & $\beta_{12}$ &  0.027 & 0.089 & 0.093 & 0.92 & -0.181 & 0.249 & 0.308 & 0.78 \\
100 & 4 & $\beta_{21}$ & -0.083 & 0.109 & 0.137 & 0.94 & -0.241 & 0.189 & 0.306 & 0.74 \\
100 & 4 & $\beta_{22}$ &  0.037 & 0.084 & 0.092 & 0.96 & -0.157 & 0.244 & 0.290 & 0.84 \\
100 & 4 & $\beta_{01}$ & -0.089 & 0.202 & 0.220 & 0.92 & -0.332 & 0.206 & 0.391 & 0.72 \\
100 & 4 & $\beta_{02}$ &  0.108 & 0.142 & 0.178 & 0.82 & -0.039 & 0.231 & 0.234 & 0.78 \\
300 & 2 & $\nu$       & -0.820 & 0.394 & 0.910 & 0.50 & 1.597 & 0.051 & 1.597 & 0.10 \\
300 & 2 & $\beta_{11}$ & -0.104 & 0.089 & 0.137 & 0.78 & -0.280 & 0.224 & 0.359 & 0.58 \\
300 & 2 & $\beta_{12}$ &  0.052 & 0.066 & 0.084 & 0.79 & -0.133 & 0.240 & 0.274 & 0.68 \\
300 & 2 & $\beta_{21}$ & -0.085 & 0.083 & 0.118 & 0.84 & -0.275 & 0.228 & 0.357 & 0.60 \\
300 & 2 & $\beta_{22}$ &  0.042 & 0.069 & 0.081 & 0.86 & -0.127 & 0.235 & 0.268 & 0.74 \\
300 & 2 & $\beta_{01}$ & -0.068 & 0.108 & 0.128 & 0.95 & -0.438 & 0.335 & 0.552 & 0.62 \\
300 & 2 & $\beta_{02}$ &  0.001 & 0.103 & 0.103 & 0.92 & -0.032 & 0.253 & 0.255 & 0.88 \\
300 & 4 & $\nu$       &  0.187 & 0.489 & 0.523 & 0.98 & 3.574 & 0.064 & 3.574 & 0.20 \\
300 & 4 & $\beta_{11}$ & -0.078 & 0.078 & 0.111 & 0.85 & -0.252 & 0.205 & 0.321 & 0.64 \\
300 & 4 & $\beta_{12}$ &  0.034 & 0.056 & 0.065 & 0.86 & -0.128 & 0.254 & 0.284 & 0.74 \\
300 & 4 & $\beta_{21}$ & -0.074 & 0.069 & 0.102 & 0.86 & -0.288 & 0.225 & 0.352 & 0.62 \\
300 & 4 & $\beta_{22}$ &  0.039 & 0.062 & 0.073 & 0.88 & -0.136 & 0.246 & 0.280 & 0.74 \\
300 & 4 & $\beta_{01}$ & -0.012 & 0.115 & 0.116 & 0.98 & -0.453 & 0.309 & 0.544 & 0.60 \\
300 & 4 & $\beta_{02}$ & -0.004 & 0.093 & 0.093 & 0.93 & -0.036 & 0.250 & 0.253 & 0.88 \\
\bottomrule
\end{tabular*}
\begin{tablenotes}[flushleft]\footnotesize
\item Bias: mean(estimate$-$true); SD: empirical standard deviation; RMSE: root mean squared error; CP: nominal 95\% coverage probability.
\end{tablenotes}
\end{threeparttable}
\end{table*}
\FloatBarrier

\begin{table*}[!t]
\centering
\begin{threeparttable}
\caption{Simulation diagnostics: Gelman--Rubin diagnostic ($\hat{R}$) and effective sample size (ESS).}
\label{tab:diagnostics}
\renewcommand{\arraystretch}{1.25}
\begin{tabular*}{\textwidth}{@{\extracolsep\fill}lccc@{}}
\toprule
\textbf{Parameters} & $\boldsymbol{\hat{R}}^{\dagger}$ & $\mathrm{ESS}^{\dagger}$ & $\mathrm{ESS}(\%)^{\dagger}$ \\
\midrule
$\nu$        & 1.00 & 1680 & 42.0\% \\
$\beta_{11}$ & 1.00 & 2769 & 69.2\% \\
$\beta_{12}$ & 1.00 & 2773 & 69.3\% \\
$\beta_{21}$ & 1.00 & 2729 & 68.2\% \\
$\beta_{22}$ & 1.00 & 2577 & 64.4\% \\
$\beta_{01}$ & 1.00 & 2646 & 66.1\% \\
$\beta_{02}$ & 1.00 & 2566 & 64.1\% \\
\bottomrule
\end{tabular*}
\begin{tablenotes}[flushleft]\footnotesize
\item[$\dagger$] $\hat{R}$ is the potential scale reduction factor and ESS is the effective sample size. Total post–burn-in draws $=4{,}000$;  $\hat{R}\le 1.01$ and ESS$\ge10\%$ to indicate good convergence quality.
\end{tablenotes}
\end{threeparttable}
\end{table*}
\FloatBarrier

\begin{table*}[!t]
\centering
\begin{threeparttable}
\caption{Frailty parameter prior sensitivity across two scenarios.}
\label{tab:prior_sensitivity_four}
\renewcommand{\arraystretch}{1.2}
\begin{tabular*}{\textwidth}{@{\extracolsep\fill}llrrrr@{}}
\toprule
Scenario & Param & Bias (G) & Bias (LN) & Diff\textsuperscript{*} & Diff $\le 0.05$ \\
\midrule
$n{=}100$, $\gamma{=}1.1$, $\nu{=}2$ & $\nu$         & -0.223 & -0.263 & 0.040 & Yes \\
                                     & $\beta_{11}$  & -0.079 & -0.078 & 0.001 & Yes \\
                                     & $\beta_{12}$  &  0.006 &  0.007 & 0.001 & Yes \\
                                     & $\beta_{21}$  & -0.077 & -0.076 & 0.001 & Yes \\
                                     & $\beta_{22}$  &  0.005 &  0.006 & 0.001 & Yes \\
                                     & $\beta_{01}$  & -0.074 & -0.073 & 0.001 & Yes \\
                                     & $\beta_{02}$  & -0.001 & -0.002 & 0.001 & Yes \\
\addlinespace
$n{=}300$, $\gamma{=}1.1$, $\nu{=}2$ & $\nu$         & -0.381 & -0.397 & 0.016 & Yes \\
                                     & $\beta_{11}$  & -0.078 & -0.078 & 0.001 & Yes \\
                                     & $\beta_{12}$  &  0.001 &  0.002 & 0.001 & Yes \\
                                     & $\beta_{21}$  & -0.078 & -0.077 & 0.001 & Yes \\
                                     & $\beta_{22}$  & -0.000 &  0.001 & 0.001 & Yes \\
                                     & $\beta_{01}$  & -0.093 & -0.093 & 0.001 & Yes \\
                                     & $\beta_{02}$  &  0.000 & -0.001 & 0.001 & Yes \\
\bottomrule
\end{tabular*}
\begin{tablenotes}[flushleft]\footnotesize
\item[\textsuperscript{*}] Diff $\equiv \big|\mathrm{Bias}_{\mathrm{Gamma}}-\mathrm{Bias}_{\mathrm{LogNormal}}\big|$; G: Gamma distribution; LN: Log Normal distribution.
\end{tablenotes}
\end{threeparttable}
\end{table*}
\FloatBarrier

\begin{table*}[!t]
\centering
\begin{threeparttable}
\caption{$R_k$ (initialization sensitivity) across two selected scenarios.}
\label{tab:Rk_two_scenarios}
\renewcommand{\arraystretch}{1.2}
\begin{tabular*}{\textwidth}{@{\extracolsep\fill}llrr@{}}
\toprule
\textbf{Scenario} & \textbf{Param} & \textbf{$R_k^\dagger$ (Bayesian)} & \textbf{$R_k^\dagger$ (Frequentist)} \\
\midrule
$n{=}100$, $\gamma{=}0.9$, $\nu{=}4$ & $\nu$        & 0.00017 & 0.46093 \\
                                     & $\beta_{11}$ & 0.00040 & 3.02383 \\
                                     & $\beta_{12}$ & 0.00003 & 0.42940 \\
                                     & $\beta_{21}$ & 0.00027 & 3.02029 \\
                                     & $\beta_{22}$ & 0.00008 & 0.41315 \\
                                     & $\beta_{01}$ & 0.00018 & 3.01996 \\
                                     & $\beta_{02}$ & 0.00027 & 0.37885 \\
\addlinespace
$n{=}300$, $\gamma{=}0.9$, $\nu{=}4$ & $\nu$        & 0.00039 & 0.43547 \\
                                     & $\beta_{11}$ & 0.00030 & 4.21679 \\
                                     & $\beta_{12}$ & 0.00013 & 0.45093 \\
                                     & $\beta_{21}$ & 0.00006 & 3.79733 \\
                                     & $\beta_{22}$ & 0.00019 & 0.39949 \\
                                     & $\beta_{01}$ & 0.00018 & 3.69231 \\
                                     & $\beta_{02}$ & 0.00015 & 0.35781 \\
\bottomrule
\end{tabular*}
\begin{tablenotes}[flushleft]\footnotesize \item[$\dagger$] $R_k \,=\, \big|\mathrm{Estimate}^{(I_1)}-\mathrm{Estimate}^{(I_2)}\big|$ for each parameter and method, where $\mathrm{Estimate}^{(I_1)}$ and $\mathrm{Estimate}^{(I_2)}$ is the parameter estimates under initialization $I_1 = 4$ and $I_2 = 1$ respectively. \end{tablenotes} \end{threeparttable}
\end{table*}
\FloatBarrier

\begin{table*}[!ht]
\centering
\caption{Computation time (seconds) per replicated dataset for each simulation scenario$^{\ddagger}$.}
\label{tab:runtime}
\renewcommand{\arraystretch}{1.2}
\setlength{\tabcolsep}{6pt}
\begin{tabular*}{\textwidth}{@{\extracolsep\fill}ccccccccc}
\toprule
\multicolumn{3}{c}{Scenario} &
\multicolumn{2}{c}{Frequentist} &
\multicolumn{2}{c}{Bayesian} &
\multicolumn{2}{c}{Runtime advantage} \\
\cmidrule(lr){1-3}\cmidrule(lr){4-5}\cmidrule(lr){6-7}\cmidrule(lr){8-9}
$n$ & $\gamma$ & $v$ & Mean & SD & Mean & SD & Speed-up$^{*}$ & \% Reduction$^{\dagger}$ \\
\midrule
100 & 0.9 & 2 & 301.1 & 81.3 & 74.5 & 11.6 & 4.04 & 75.3 \\
100 & 0.9 & 4 & 289.3 & 81.6 & 75.0 & 12.6 & 3.86 & 74.1 \\
100 & 1.1 & 2 & 299.9 & 70.6 & 72.7 & 10.0 & 4.13 & 75.8 \\
100 & 1.1 & 4 & 279.5 & 69.2 & 72.5 & 12.4 & 3.86 & 74.1 \\
\addlinespace
300 & 0.9 & 2 & 329.0 & 61.3 & 137.2 & 17.4 & 2.40 & 58.3 \\
300 & 0.9 & 4 & 309.6 & 54.6 & 138.3 & 16.6 & 2.24 & 55.3 \\
300 & 1.1 & 2 & 324.6 & 52.5 & 138.0 & 18.8 & 2.35 & 57.5 \\
300 & 1.1 & 4 & 304.6 & 47.5 & 136.3 & 16.7 & 2.23 & 55.2 \\
\bottomrule
\end{tabular*}
\vspace{-0.4cm}
{\footnotesize
\begin{flushleft}
$^{*}$ Speed-up: frequentist mean divided by Bayesian mean ($>1$ favors Bayesian).\\
$^{\dagger}$ \% Reduction: decrease in mean time for Bayesian relative to frequentist.\\
$^{\ddagger}$ Times are per replicated dataset based on 5000 iterations.
\end{flushleft}
}
\end{table*}
\FloatBarrier

\begin{figure}[!ht]\label{fig2}
\centering
\includegraphics[width=\columnwidth]{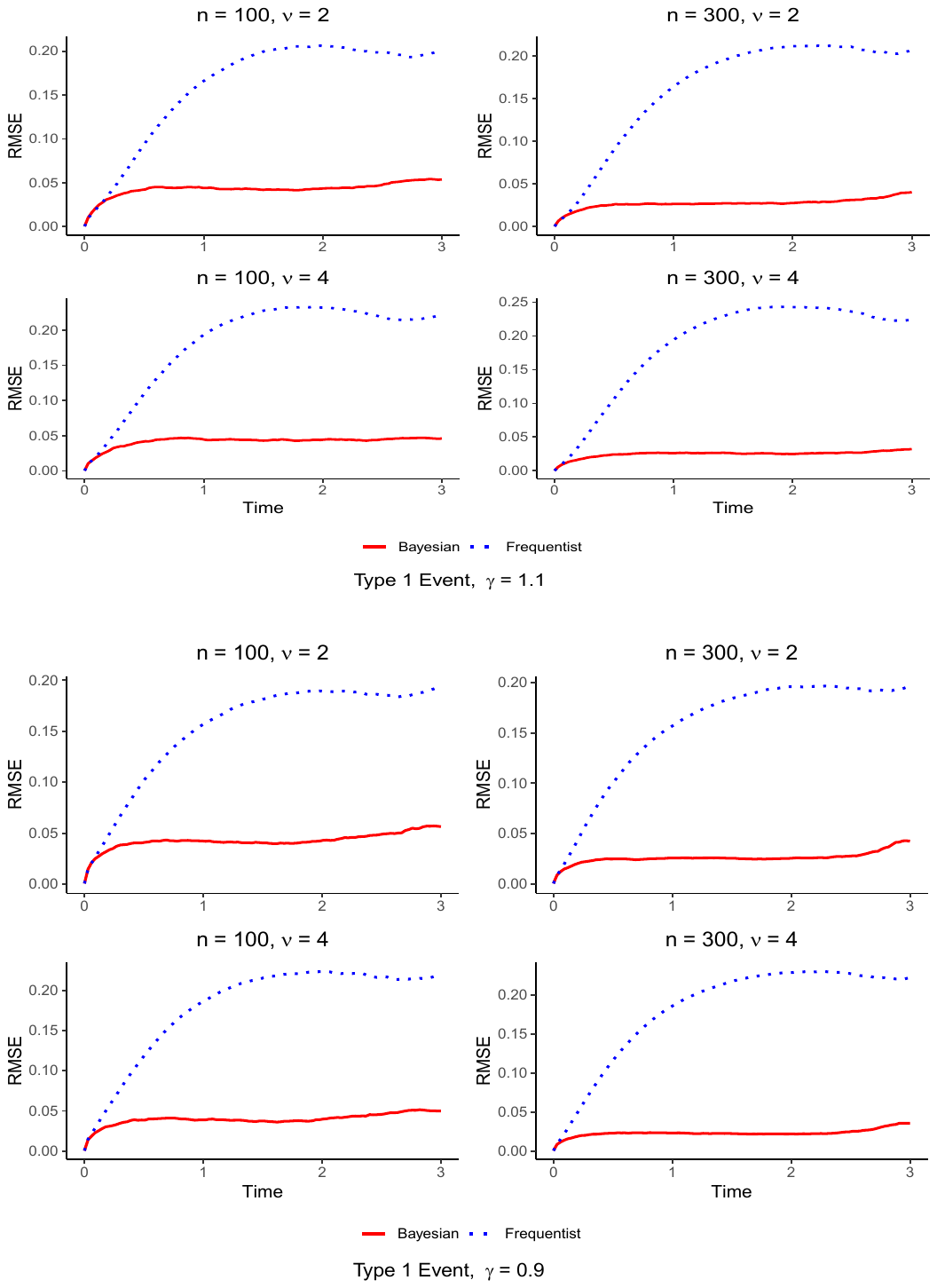}
\vspace{-0.7in}
\caption{Simulated RMSE for the baseline survival function estimator $\hat{\bar F}_1(t)$ of the gap-time distribution, comparing Bayesian and frequentist models for a type~1 event under (a) increasing failure rate ($\gamma=1.1$) and (b) decreasing failure rate ($\gamma=0.9$).}
\label{fig:Figure_2_Mithun.pdf}
\end{figure}
 \FloatBarrier 
 
\begin{figure}[!ht]\label{fig3}
\centering
\includegraphics[width=\columnwidth]{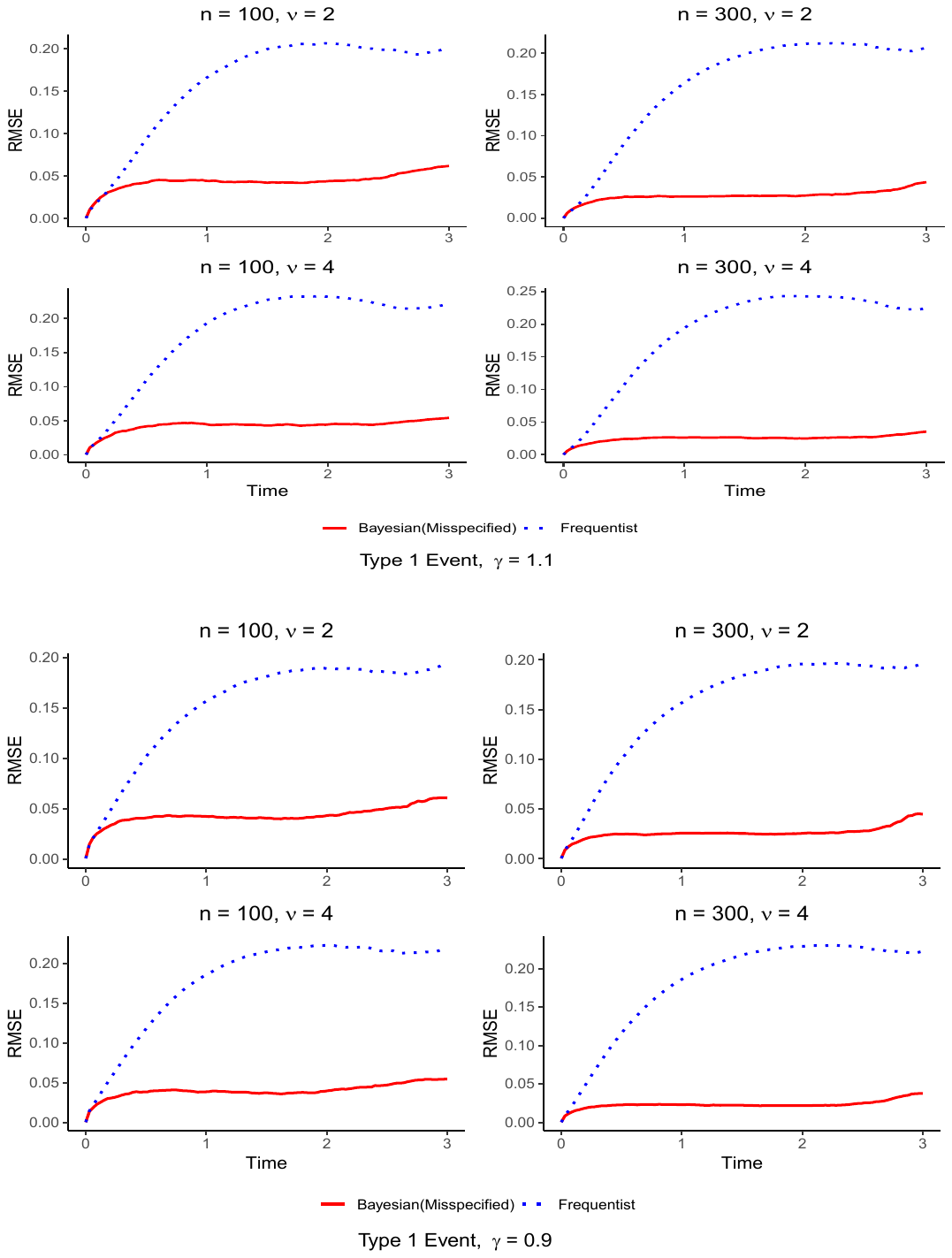}
\vspace{-0.7in}
\caption{Simulated RMSE for the baseline survival function estimator $\hat{\bar F}_1(t)$ of the gap-time distribution, comparing a Bayesian model with misspecified priors and a frequentist model for a type~1 event under (a) increasing failure rate ($\gamma=1.1$) and (b) decreasing failure rate ($\gamma=0.9$).}
\label{fig:Figure_3_Mithun.pdf}
\end{figure}
 \FloatBarrier 
 
\begin{figure}[!ht]\label{fig4}
\centering
\includegraphics[width=\columnwidth]{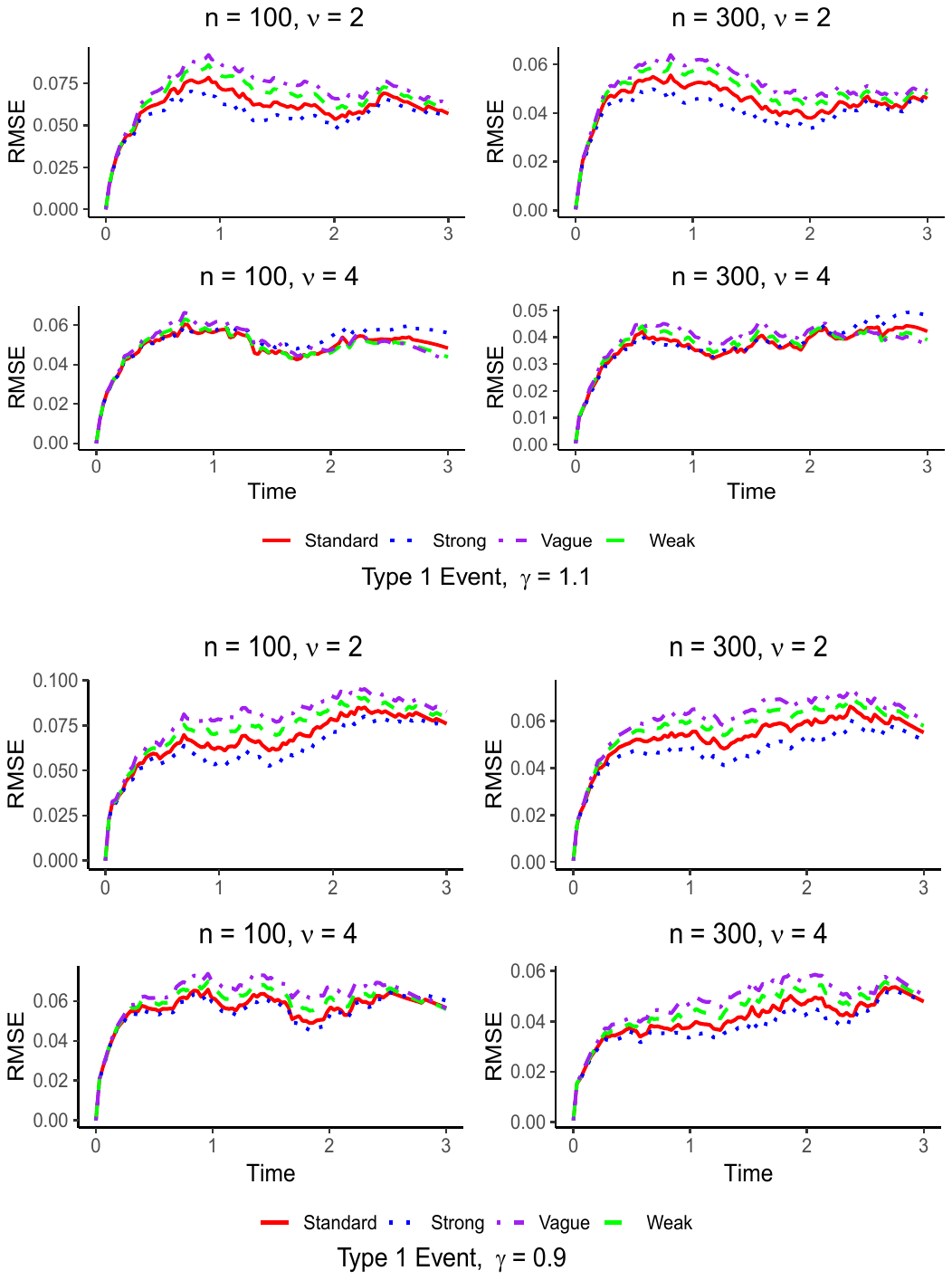}
\vspace{-0.7in}
\caption{Simulated RMSE for the baseline survival function estimator $\hat{\bar F}_1(t)$ of the gap-time distribution, comparing Bayesian and frequentist models for a type~1 event under (a) increasing failure rate ($\gamma=1.1$) and (b) decreasing failure rate ($\gamma=0.9$). (Priors: strong $\{0.5,\,0.25\}$; standard $\{1,\,0.5\}$; weak $\{2.25,\,1\}$; vague $\{9,\,2\}$ for $(\sigma^2_{\beta},\sigma^2_{\nu})$.}
\label{fig:Figure_4_Mithun.pdf}
\end{figure}
 \FloatBarrier 
 
\begin{figure}[!ht]\label{fig5}
\centering
\includegraphics[width=\columnwidth]{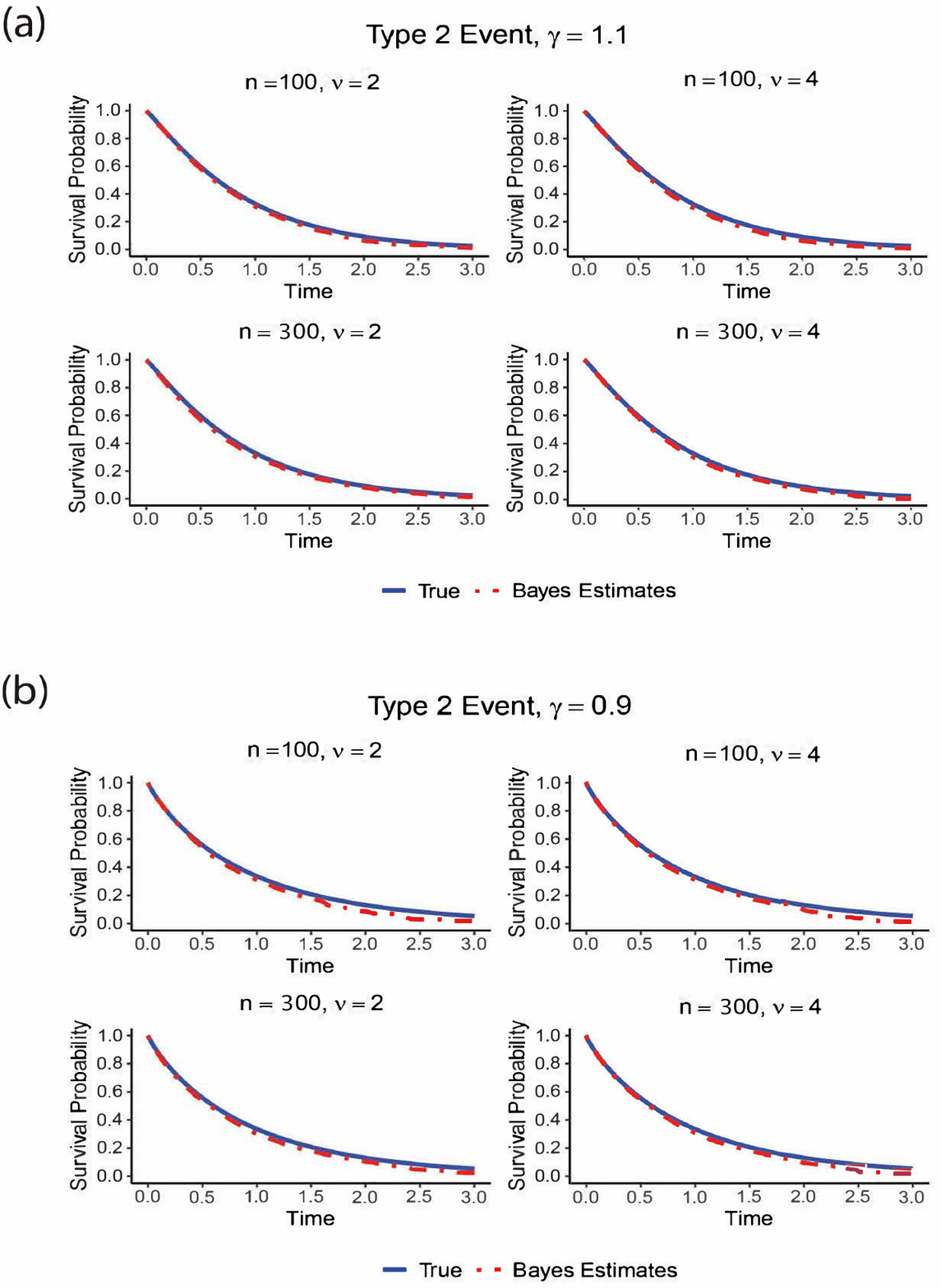}
\caption{True versus estimated baseline survival functions for type~2 events across simulation settings.}
\label{fig:Figure_5_Mithun.pdf}
\end{figure}
 \FloatBarrier 
 
Pointwise RMSE curves for the baseline survival functions favor the Bayesian estimator in nearly all settings for type 1 events, with visibly smaller errors that shrink further at \(n=300\) (Figure~\ref{fig:Figure_2_Mithun.pdf}). We observed similar results for type 2 and for the terminal event. Figure~\ref{fig:Figure_5_Mithun.pdf} shows that the Bayesian estimated survival curves for type 2 events sit almost on top of the true curves in every scenario, with gaps getting smaller at \(n=300\). The same close match holds for type 1 events and for the terminal event. Taken together, Figures~\ref{fig:Figure_2_Mithun.pdf} and \ref{fig:Figure_5_Mithun.pdf} show lower pointwise RMSE and tight truth–estimate overlays for the Bayesian model, indicating stronger predictive performance across sample sizes and hazard shapes.

Table~\ref{tab:diagnostics} shows that all chains mix well with \(\widehat R=1.00\) and large ESS (64–69\% for the \(\beta\)’s and 42\% for \(\nu\)), which supports reliable posterior summaries. Changing prior strength for the regressions and for \(\nu\) from strong to vague gives almost overlapping RMSE curves, so inference is not sensitive to how informative the priors are (Figure~\ref{fig:Figure_4_Mithun.pdf}). Replacing the Weibull working baseline with an exponential one leaves survival RMSE and the truth–estimate overlays essentially unchanged, which shows that prediction does not depend on the baseline prior form (Figure~\ref{fig:Figure_3_Mithun.pdf}). Switching the frailty prior from Gamma to a moment–matched log–normal yields bias differences no larger than \(0.05\) across the reported parameters (Table~\ref{tab:prior_sensitivity_four}). Finally, dispersed initial values have little effect on the Bayesian estimates, whereas the EM fit moves more under the same starts (Table~\ref{tab:Rk_two_scenarios}). Taken together, these checks point to stable behavior under changes in convergence, prior strength and form, baseline specification, and initialization, with results that consistently favor the Bayesian approach (Figures~\ref{fig:Figure_3_Mithun.pdf}–\ref{fig:Figure_4_Mithun.pdf}; Tables~\ref{tab:diagnostics}–\ref{tab:Rk_two_scenarios}).

Table~\ref{tab:runtime} shows that the Bayesian fit runs faster than EM in every scenario. At \(n=100\) it is about \(3.9\)–\(4.1\times\) faster, cutting the average time by \(74\%\)–\(76\%\); at \(n=300\) it is about \(2.2\)–\(2.4\times\) faster, with a \(55\%\)–\(58\%\) reduction. Time increases with sample size for both methods, but the gap stays large. The faster run time makes routine re-fits and sensitivity checks easy to do. In large studies—such as the ALLHAT trial—these savings can help complete full analyses and the usual checks on a reasonable schedule.

\section{Application}\label{sec:appli}
The ALLHAT trial was a large-scale, double-blind, randomized trial administered among 623 centers in North America from February 1994 to March 2002 \cite{antihypertensive2002major}. The trial was funded by the National Heart, Lung, and Blood Institute (NHLBI). A key objective of this trial was to compare the effectiveness of various antihypertensive medications in preventing major heart attacks in high-risk patients. The study included 33,357 hypertensive patients with at least one additional cardiovascular risk factor. Among them, 15,255 (45.73\%) received Chlorthalidone, 9048 (27.12\%) received Amlodipine, and 9054 (27.15\%) received Lisinopril. The participants were followed for an average of 4.9 years \cite{wright2005outcomes}. In this pragmatic trial, 95\% of participants completed follow-up assessments.

During the ALLHAT trial, a total of five major cardiovascular events were recorded: myocardial infarction, stroke, congestive heart failure, angina, and peripheral arterial disease. During the follow-up period, 4.0\% of participants had myocardial infarction, 3.4\% had stroke, 4.4\% had congestive heart failure, 7.1\% had angina, and 1.6\% had peripheral arterial disease. The impact of these cardiovascular events on hypertensive patients is high, suggesting the need for long-term treatment. However, clinical signs and symptoms, treatment choices, and disease development differ between events. Thus, it is crucial to distinguish them when evaluating cardiovascular risk.

Generally, myocardial infarction and stroke occur suddenly due to acute thrombotic or ischemic events. This requires immediate medical intervention, such as thrombolytic therapy, percutaneous coronary intervention (PCI), or urgent stroke management \cite{benjamin2019heart,fuster2014acute}. By contrast, prolonged cardiovascular dysfunction is associated with chronic cardiovascular diseases, including congestive heart failure, angina, and peripheral arterial disease; these are driven by sustained vascular damage, impaired regulation of blood flow, and endothelial dysfunction \cite{mh20172017,fowkes2013comparison}. These chronic conditions are managed through lifestyle modification, medical treatments (e.g., beta-blockers, ACE inhibitors, statins), and sometimes revascularization procedures \cite{Roffi2016,GerhardHerman2017}. Taking these disease patterns into account, we classified cardiovascular events as acute cardiovascular events (ACEs) (myocardial infarction and stroke) and chronic cardiovascular events (CCEs) (congestive heart failure, angina, and peripheral arterial disease). This classification clarifies immediate cardiovascular threats versus long-term disease burden and supports precision risk modeling and treatment stratification \cite{Go2013,Townsend2016}.

\begin{table*}[!ht]
\centering
\begin{threeparttable}
\caption{Hazard ratios (HR) and  $95\%$ credible intervals}
\label{tab:estimates}
\renewcommand{\arraystretch}{1.25}
\setlength{\tabcolsep}{5pt}
\begin{tabular*}{\textwidth}{@{\extracolsep\fill}llrrrrr@{}}
\toprule
\textbf{Event} & \textbf{Variable} & \textbf{Estimate} & \textbf{SE} & \textbf{Lower CI} & \textbf{Upper CI} & \textbf{HR}$^\ast$ \\
\midrule
ACE  & Amlodipine          & 0.57 & 0.05 & 0.49 & 0.68 & 1.77 \\
     & Lisinopril          & 0.83 & 0.06 & 0.71 & 0.92 & 2.30 \\
     & Race (Black)        & 1.42 & 0.05 & 1.34 & 1.53 & 4.15 \\
     & Age$^{\dagger}$     & 1.45 & 0.04 & 1.39 & 1.54 & 4.28 \\
\addlinespace
CCE  & Amlodipine          & 0.65 & 0.04 & 0.57 & 0.72 & 1.92 \\
     & Lisinopril          & 0.79 & 0.04 & 0.71 & 0.86 & 2.20 \\
     & Race (Black)        & 1.52 & 0.06 & 1.41 & 1.62 & 4.56 \\
     & Age$^{\dagger}$     & 1.34 & 0.04 & 1.29 & 1.42 & 3.82 \\
\addlinespace
Death& Amlodipine          & 0.88 & 0.04 & 0.81 & 0.95 & 2.41 \\
     & Lisinopril          & 1.09 & 0.05 & 0.98 & 1.17 & 2.98 \\
     & Race (Black)        & 1.91 & 0.06 & 1.80 & 2.02 & 6.78 \\
     & Age$^{\dagger}$     & 1.89 & 0.03 & 1.83 & 1.96 & 6.59 \\
\addlinespace
---  & Frailty ($\nu$)     & 0.12 & 0.002& 0.11 & 0.13 & \textemdash \\
\bottomrule
\end{tabular*}
\begin{tablenotes}[flushleft]\footnotesize
\item[$\dagger$] Age is standardized to mean $0$ and variance $1$. Chlorthalidone is the reference treatment.
\item[$\ast$] HR $=\exp(\text{Estimate})$ for regression effects on the log-hazard scale.
\end{tablenotes}
\end{threeparttable}
\end{table*}
 \FloatBarrier   
 
\begin{figure*}[!ht]
  \centering
  \includegraphics[
    width=\linewidth,
    height=0.86\textheight, 
    keepaspectratio
  ]{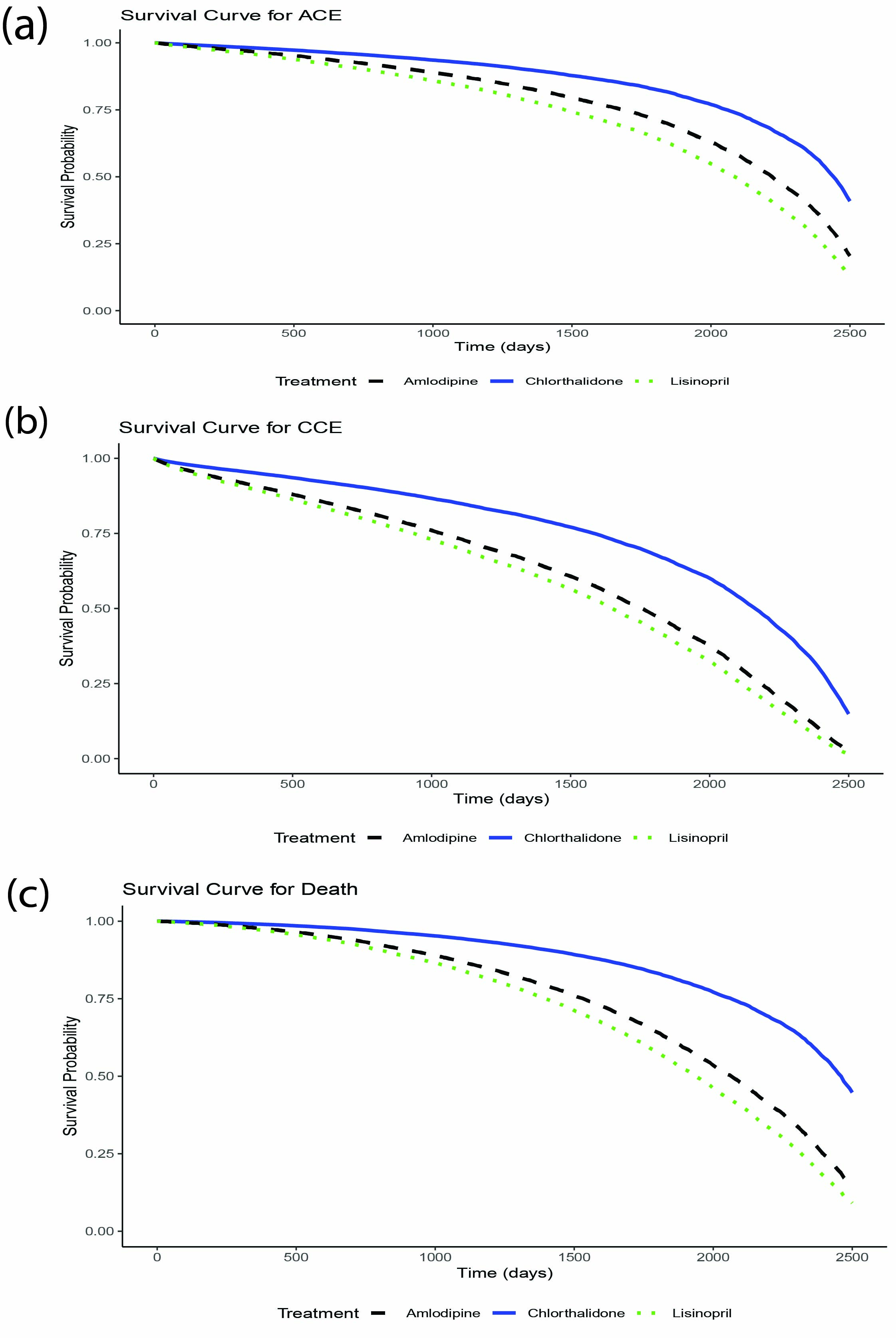}
  \caption{Treatment-specific survival probabilities for (a) acute cardiovascular events,
  (b) chronic cardiovascular events, and (c) death, comparing Chlorthalidone, Amlodipine,
  and Lisinopril.}
  \label{fig:survival-treatments}
\end{figure*}

Among the 33,357 participants, 1423 (2.0\%) experienced ACE, 2890 (4.0\%) experienced CCE, and 931 (1.3\%) had both ACE and CCE events during the follow-up period. Moreover, 4929 (14.8\%) of hypertensive patients died from an all-cause event, underscoring the substantial burden of cardiovascular disease among hypertensive patients. Multiple cardiovascular incidents can accelerate the onset of disease toward this high mortality. This suggests an interrelation between recurrent and terminal events and underscores the need for a comprehensive multitype recurrent event framework incorporating frailty terms.

In Table~\ref{tab:estimates}, a positive coefficient was found for Amlodipine (ref.\ = Chlorthalidone) for ACE, which is statistically significant. Compared with Chlorthalidone, the effect of Lisinopril was statistically significant, with an estimated association corresponding to HR = 2.30 (95\% CI: [2.03, 2.51]), indicating more than a twofold increase in hazard. The association between ACE and Black race was statistically significant with a high hazard ratio (HR = 4.15, 95\% CI: [3.81, 4.62]). ACE also shows a statistically significant relationship with age (HR = 4.28). For CCE, both Amlodipine and Lisinopril showed strong and statistically significant effects, indicating 92\% and 120\% higher hazards compared with Chlorthalidone. Black race had a 4.56-fold higher risk of CCE (HR = 4.56), whereas age had a 3.82-fold higher risk (HR = 3.82). Both race and age show statistically significant associations with CCE. In comparison with Amlodipine and Lisinopril, Chlorthalidone had a significantly lower death hazard, with reciprocal hazard ratios of 0.41 (1/2.41) and 0.34 (1/2.98), respectively, suggesting 59\% and 66\% reductions in mortality. As with ACE and CCE, Black race was a significant predictor of death, with a substantially higher risk (HR = 6.78, 95\% CI: [6.05, 7.54]) than for non-Black participants. Age was also a statistically significant predictor, with HR = 6.59 (95\% CI: [6.23, 7.10]). Based on the estimate of the frailty parameter of 0.12 (95\% CI: [0.11, 0.13]), there is high unobserved heterogeneity and positive dependence within subjects across event types. To estimate the parameters using MCMC, we used 2000 iterations, 1000 burn-in, and a thinning of 5.

Based on the three treatment groups, Figure~\ref{fig:survival-treatments}(a–c) illustrates the survival probabilities for ACE, CCE, and death. As seen in Figure~\ref{fig:survival-treatments}(a), the survival curve for Chlorthalidone (the reference treatment) shows the highest survival probability. In contrast, Amlodipine and Lisinopril exhibit nearly identical risk profiles, with clearly lower survival probabilities than the reference. Chlorthalidone demonstrated the highest survival rate in the CCE cohort (Figure~\ref{fig:survival-treatments}(b)). The two treatments (Amlodipine and Lisinopril) exhibited similarly low survival rates, suggesting increased risk relative to the reference. In the case of death (Figure~\ref{fig:survival-treatments}(c)), the reference group maintained the highest survival, followed by Amlodipine, with Lisinopril showing the lowest survival probability. Based on this pattern, for each event type the reference drug Chlorthalidone has the highest survival, followed by Amlodipine and then Lisinopril.

\section{Discussion}\label{sec:diss}
We propose a semiparametric Bayesian joint frailty model within a gap-time framework for analyzing recurrent events as well as terminal events. The shared frailty structure manages subject-level dependence and heterogeneous event risks without compromising computational efficiency. We derive a closed-form expression for the cumulative hazard function, which ensures precise estimation and simplifies comparisons across event types. In addition, the model places a gamma–process prior on the baseline cumulative hazard function. This prior yields smooth, nonparametric estimation while maintaining accuracy and computational efficiency, supporting robust inference under complex hazard patterns. The model is therefore both theoretically sound and practically efficient. Overall, it is particularly well suited to biomedical data characterized by heterogeneous risks and sparse occurrences.

In comparison with a frequentist EM-based version of a joint frailty approach, our Bayesian method shows better performance. Across multiple settings (varying sample size, event rates, and frailty values), the model consistently produced more accurate estimates of cumulative hazards than the frequentist comparator. Moreover, the method remained competitive even when baseline-hazard priors were misspecified, underscoring the model’s ability to adapt to prior assumptions. A prior sensitivity analysis further indicated that performance was remarkably stable across hyperparameter choices, suggesting strong robustness built into the model structure.

We applied the model to the ALLHAT dataset to analyze five cardiovascular outcomes classified into acute cardiovascular events (ACE) and chronic cardiovascular events (CCE), along with all-cause mortality as a terminal event. The joint specification accounts for hidden heterogeneity linking the recurrent processes (ACE and CCE) with the terminal event (death). The fitted model highlights meaningful differences in risk patterns across ACE, CCE, and death for both treatments and predictors: Lisinopril and Amlodipine exhibit higher hazards for death than for ACE or CCE, and—consistent with the trial’s reference comparison—the Chlorthalidone arm shows the most favorable profile overall. Race (Black vs.\ non-Black) and age are statistically significant predictors across event types, with the largest hazard ratios observed for death. These findings align with the original clinical results reported in the JAMA publications \cite{antihypertensive2002major,wright2005outcomes}.

Despite the strengths of the model, it assumes independent subject-level frailties and noninformative censoring, which may limit generalizability in some settings. Future research could relax these assumptions to capture interdependencies between event types and to accommodate dependent censoring mechanisms. Nevertheless, the comprehensive joint framework, coupled with the Bayesian formulation, makes the approach valuable for applied researchers seeking to understand complex event dynamics and their determinants. Extensions could also incorporate dynamic effects of past events on subsequent recurrences and/or the terminal event.

 \section*{Declarations}

\subsection*{Funding}
No funding was received to assist with the preparation of this manuscript. 

\subsection*{Competing Interests}
The authors declare that they have no competing interests. 

 \subsection*{Acknowledgments}
 We thank the National Heart, Lung, and Blood Institute (NIH) for providing access to data from the Antihypertensive and Lipid-Lowering Treatment to Prevent Heart Attack Trial (ALLHAT).


\begin{thebibliography}{99}

\bibitem{ravani2013temporal}
Ravani, P., Gillespie, B. W., Quinn, R. R., MacRae, J., Manns, B., Mendelssohn, D., Tonelli, M., Hemmelgarn, B., James, M., Pannu, N., \textit{et al.} (2013).
Temporal risk profile for infectious and noninfectious complications of hemodialysis access.
\textit{Journal of the American Society of Nephrology}, 24(10), 1668--1677.
https://doi.org/10.1681/ASN.2012121234

\bibitem{tuli2000risk}
Tuli, S., Drake, J., Lawless, J., Wigg, M., \& Lamberti-Pasculli, M. (2000).
Risk factors for repeated cerebrospinal shunt failures in pediatric patients with hydrocephalus.
\textit{Journal of Neurosurgery}, 92(1), 31--38.
https://doi.org/10.3171/jns.2000.92.1.0031

\bibitem{byar1980veterans}
Byar, D. P. (1980).
The Veterans Administration study of chemoprophylaxis for recurrent stage I bladder tumours: Comparisons of placebo, pyridoxine and topical thiotepa.
In \textit{Bladder Tumors and Other Topics in Urological Oncology} (pp. 363--370). Springer.

\bibitem{andersen1982cox}
Andersen, P. K., \& Gill, R. D. (1982).
Cox’s regression model for counting processes: A large sample study.
\textit{The Annals of Statistics}, 10(4), 1100--1120.
https://doi.org/10.1214/aos/1176345976

\bibitem{andersen2012statistical}
Andersen, P. K., Borgan, {\O}., Gill, R. D., \& Keiding, N. (2012).
\textit{Statistical Models Based on Counting Processes}. Springer.
https://doi.org/10.1007/978-1-4612-4348-9

\bibitem{pepe1993some}
Pepe, M. S., \& Cai, J. (1993).
Some graphical displays and marginal regression analyses for recurrent failure times and time-dependent covariates.
\textit{Journal of the American Statistical Association}, 88(423), 811--820.
https://doi.org/10.1080/01621459.1993.10476346

\bibitem{lawless1995some}
Lawless, J. F., \& Nadeau, C. (1995).
Some simple robust methods for the analysis of recurrent events.
\textit{Technometrics}, 37(2), 158--168.
https://doi.org/10.1080/00401706.1995.10484302

\bibitem{lin2000semiparametric}
Lin, D. Y., Wei, L.-J., Yang, I., \& Ying, Z. (2000).
Semiparametric regression for the mean and rate functions of recurrent events.
\textit{Journal of the Royal Statistical Society: Series B}, 62(4), 711--730.
https://doi.org/10.1111/1467-9868.00264

\bibitem{pena2001nonparametric}
Pe\~na, E. A., Strawderman, R. L., \& Hollander, M. (2001).
Nonparametric estimation with recurrent event data.
\textit{Journal of the American Statistical Association}, 96(456), 1299--1315.
https://doi.org/10.1198/016214501753381922

\bibitem{pena2007semiparametric}
Pe\~na, E. A., Slate, E. H., \& Gonz\'alez, J. R. (2007).
Semiparametric inference for a general class of models for recurrent events.
\textit{Journal of Statistical Planning and Inference}, 137(6), 1727--1747.
https://doi.org/10.1016/j.jspi.2006.05.014

\bibitem{rahman2014nonparametric}
Rahman, A. K. M. F., Lynch, J. D., \& Pe\~na, E. A. (2014).
Nonparametric Bayes estimation of gap-time distribution with recurrent event data.
\textit{Journal of Nonparametric Statistics}, 26(3), 575--598.
https://doi.org/10.1080/10485252.2014.910698

\bibitem{zhangsheng2011joint}
Yu, Z., \& Liu, L. (2011).
A joint model of recurrent events and a terminal event with a nonparametric covariate function.
\textit{Statistics in Medicine}, 30(22), 2683--2695.
https://doi.org/10.1002/sim.4263

\bibitem{yu2014joint}
Yu, Z., Liu, L., Bravata, D. M., \& Williams, L. S. (2014).
Joint model of recurrent events and a terminal event with time-varying coefficients.
\textit{Biometrical Journal}, 56(2), 183--197.
https://doi.org/10.1002/bimj.201200240

\bibitem{ghosh2000nonparametric}
Ghosh, D., \& Lin, D. Y. (2000).
Nonparametric analysis of recurrent events and death.
\textit{Biometrics}, 56(2), 554--562.
https://doi.org/10.1111/j.0006-341X.2000.00554.x

\bibitem{li2019bayesian}
Li, Z., Chinchilli, V. M., \& Wang, M. (2019).
A Bayesian joint model of recurrent events and a terminal event.
\textit{Biometrical Journal}, 61(1), 187--202.
https://doi.org/10.1002/bimj.201700302

\bibitem{cook2007statistical}
Cook, R. J., \& Lawless, J. F. (2007).
\textit{The Statistical Analysis of Recurrent Events}. Springer.
https://doi.org/10.1007/978-0-387-69810-6

\bibitem{huang2004joint}
Huang, C.-Y., \& Wang, M.-C. (2004).
Joint modeling and estimation for recurrent event processes and failure time data.
\textit{Journal of the American Statistical Association}, 99(468), 1153--1165.
https://doi.org/10.1198/016214504000001116

\bibitem{huang2007joint}
Huang, X., \& Liu, L. (2007).
A joint frailty model for survival and gap times between recurrent events.
\textit{Biometrics}, 63(2), 389--397.
https://doi.org/10.1111/j.1541-0420.2006.00719.x

\bibitem{rondeau2007joint}
Rondeau, V., Mathoulin-Pelissier, S., Jacqmin-Gadda, H., Brouste, V., \& Soubeyran, P. (2007).
Joint frailty models for recurring events and death using maximum penalized likelihood estimation: Application on cancer events.
\textit{Biostatistics}, 8(4), 708--721.
https://doi.org/10.1093/biostatistics/kxl043

\bibitem{hougaard2000analysis}
Hougaard, P. (2000).
\textit{Analysis of Multivariate Survival Data}. Springer.
https://doi.org/10.1007/978-1-4612-1304-8

\bibitem{ripatti2000estimation}
Ripatti, S., \& Palmgren, J. (2000).
Estimation of multivariate frailty models using penalized partial likelihood.
\textit{Biometrics}, 56(4), 1016--1022.
https://doi.org/10.1111/j.0006-341X.2000.01016.x

\bibitem{duchateau2008frailty}
Duchateau, L., \& Janssen, P. (2008).
\textit{The Frailty Model}. Springer.
(ebook DOI: 10.1007/978-0-387-72835-3)

\bibitem{mclachlan2008em}
McLachlan, G. J., \& Krishnan, T. (2008).
\textit{The EM Algorithm and Extensions} (2nd ed.). Wiley.
(ebook DOI: 10.1002/9780470191613)

\bibitem{kober1995clinical}
K{\o}ber, L., Torp-Pedersen, C., Carlsen, J. E., Bagger, H., Eliasen, P., Lyngborg, K., Videb{\ae}k, J., Cole, D. S., Auclert, L., Pauly, N. C., \textit{et al.} (1995).
A clinical trial of the angiotensin-converting--enzyme inhibitor trandolapril in patients with left ventricular dysfunction after myocardial infarction.
\textit{New England Journal of Medicine}, 333(25), 1670--1676.
https://doi.org/10.1056/NEJM199512213332503

\bibitem{kannel1979diabetes}
Kannel, W. B., \& McGee, D. L. (1979).
Diabetes and cardiovascular disease: The Framingham study.
\textit{JAMA}, 241(19), 2035--2038.
https://doi.org/10.1001/jama.241.19.2035

\bibitem{writing2002risks}
Writing Group for the Women’s Health Initiative Investigators. (2002).
Risks and benefits of estrogen plus progestin in healthy postmenopausal women: Principal results from the Women’s Health Initiative randomized controlled trial.
\textit{JAMA}, 288(3), 321--333.
https://doi.org/10.1001/jama.288.3.321

\bibitem{aric1989atherosclerosis}
The ARIC Investigators. (1989).
The Atherosclerosis Risk in Communities (ARIC) study: Design and objectives.
\textit{American Journal of Epidemiology}, 129(4), 687--702.
https://doi.org/10.1093/oxfordjournals.aje.a115184

\bibitem{ghosh2023dynamic}
Ghosh, A., Chan, W., Younes, N., \& Davis, B. R. (2023).
A dynamic risk model for multitype recurrent events.
\textit{American Journal of Epidemiology}, 192(4), 621--631.
https://doi.org/10.1093/aje/kwac229

\bibitem{lin2017bayesian}
Lin, L.-A., Luo, S., Chen, B. E., \& Davis, B. R. (2017).
Bayesian analysis of multi-type recurrent events and dependent termination with nonparametric covariate functions.
\textit{Statistical Methods in Medical Research}, 26(6), 2869--2884.
https://doi.org/10.1177/0962280215597584

\bibitem{liu2015dynamic}
Liu, P., \& Pe\~na, E. A. (2015).
Dynamic modeling \& analysis of recurrent competing risks and a terminal event.
In \textit{Proceedings of the 16th Applied Stochastic Models and Data Analysis (ASMDA 2015) International Conference}, University of Piraeus, Greece, June 30--July 4, 2015, Book 1, p.~111.

\bibitem{ibrahim2001}
Ibrahim, J. G., Chen, M.-H., \& Sinha, D. (2001).
\textit{Bayesian Survival Analysis}. Springer.
https://doi.org/10.1007/978-1-4757-3447-8

\bibitem{jachno2021impact}
Jachno, K., Heritier, S., \& Wolfe, R. (2021).
Impact of a non-constant baseline hazard on detection of time-dependent treatment effects: A simulation study.
\textit{BMC Medical Research Methodology}, 21(1), 177.
https://doi.org/10.1186/s12874-021-01365-y

\bibitem{pena1993small}
Pe\~na, E. A., \& Rohatgi, V. K. (1993).
Small sample and efficiency results for the Nelson--Aalen estimator.
\textit{Journal of Statistical Planning and Inference}, 37(2), 193--202.
https://doi.org/10.1016/0378-3758(93)90123-4

\bibitem{prentice1981regression}
Prentice, R. L., Williams, B. J., \& Peterson, A. V. (1981).
On the regression analysis of multivariate failure time data.
\textit{Biometrika}, 68(2), 373--379.
https://doi.org/10.1093/biomet/68.2.373

\bibitem{amorim2015modelling}
Amorim, L. D. A. F., \& Cai, J. (2015).
Modelling recurrent events: A tutorial for analysis in epidemiology.
\textit{International Journal of Epidemiology}, 44(1), 324--333.
https://doi.org/10.1093/ije/dyu222

\bibitem{kelly2000survival}
Kelly, P. J., \& Lim, L. L.-Y. (2000).
Survival analysis for recurrent event data: An application to childhood infectious diseases.
\textit{Statistics in Medicine}, 19(1), 13--33.
https://doi.org/10.1002/(SICI)1097-0258(20000115)19:1<13::AID-SIM275>3.0.CO;2-G

\bibitem{wang1999nonparametric}
Wang, M.-C., \& Chang, S.-H. (1999).
Nonparametric estimation of a recurrent survival function.
\textit{Journal of the American Statistical Association}, 94(445), 146--153.
https://doi.org/10.1080/01621459.1999.10474132

\bibitem{bailey2015estimation}
Bailey, L., Weaver, F. M., Chin, A. S., \& Carbone, L. D. (2015).
Estimation of a recurrent event gap-time distribution: An application to morbidity outcomes following lower extremity fracture in veterans with spinal cord injury.
\textit{Health Services and Outcomes Research Methodology}, 15, 1--22.
https://doi.org/10.1007/s10742-014-0122-8

\bibitem{li2021bayesian}
Li, Y., Seo, S., \& Lee, K. H. (2021).
Bayesian survival analysis using gamma processes with adaptive time partition.
\textit{Journal of Statistical Computation and Simulation}, 91(14), 2937--2952.
https://doi.org/10.1080/00949655.2021.1896749

\bibitem{davis1996rationale}
Davis, B. R., Cutler, J. A., Gordon, D. J., Furberg, C. D., Wright, J. T., Jr., Cushman, W. C., Grimm, R. H., LaRosa, J., Whelton, P. K., Perry, H. M., \textit{et al.} (1996).
Rationale and design for the Antihypertensive and Lipid Lowering Treatment to Prevent Heart Attack Trial (ALLHAT).
\textit{American Journal of Hypertension}, 9(4), 342--360.
https://doi.org/10.1016/0895-7061(96)00037-4

\bibitem{jacod1975multivariate}
Jacod, J. (1975).
Multivariate point processes: predictable projection, Radon--Nikodym derivatives, representation of martingales.
\textit{Zeitschrift f\"ur Wahrscheinlichkeitstheorie und verwandte Gebiete}, \textbf{31}(3), 235--253.

\bibitem{pena2000weak}
Pe\~na, E. A., Strawderman, R. L., \& Hollander, M. (2000).
A weak convergence result relevant in recurrent and renewal models.
In \textit{Recent Advances in Reliability Theory: Methodology, Practice, and Inference} (pp.~493--514). Springer.

\bibitem{kalbfleisch1978non}
Kalbfleisch, J. D. (1978).
Non-parametric Bayesian analysis of survival time data.
\textit{Journal of the Royal Statistical Society: Series B (Methodological)}, \textbf{40}(2), 214--221.

\bibitem{hougaard1995}
Hougaard, P. (1995).
Frailty models for survival data.
\textit{Lifetime Data Analysis}, \textbf{1}(3), 255--273.
https://doi.org/10.1007/BF00985760

\bibitem{ripatti2000}
Ripatti, S., \& Palmgren, J. (2000).
Estimation of multivariate frailty models using penalized partial likelihood.
\textit{Biometrics}, \textbf{56}(4), 1016--1022.
https://doi.org/10.1111/j.0006-341X.2000.01016.x

\bibitem{therneau2003}
Therneau, T. M., Grambsch, P. M., \& Pankratz, V. S. (2003).
Penalized survival models and frailty.
\textit{Journal of Computational and Graphical Statistics}, \textbf{12}(1), 156--175.
https://doi.org/10.1198/1061860031365

\bibitem{antihypertensive2002major}
ALLHAT Collaborative Research Group. (2002).
Major outcomes in high-risk hypertensive patients randomized to angiotensin-converting enzyme inhibitor or calcium channel blocker vs diuretic: The ALLHAT trial.
\textit{JAMA}, \textbf{288}(23), 2981--2997.
https://doi.org/10.1001/jama.288.23.2981

\bibitem{wright2005outcomes}
Wright, J. T., Jr., Dunn, J. K., Cutler, J. A., Davis, B. R., Cushman, W. C., Ford, C. E., Haywood, L. J., Leenen, F. H. H., Margolis, K. L., Papademetriou, V., \textit{et al.} (2005).
Outcomes in hypertensive Black and non-Black patients treated with chlorthalidone, amlodipine, and lisinopril.
\textit{JAMA}, \textbf{293}(13), 1595--1608.
https://doi.org/10.1001/jama.293.13.1595

\bibitem{benjamin2019heart}
Benjamin, E. J., Muntner, P., Alonso, A., Bittencourt, M. S., Callaway, C. W., Carson, A. P., Chamberlain, A. M., Chang, A. R., Cheng, S., Das, S. R., \textit{et al.} (2019).
Heart disease and stroke statistics---2019 update: A report from the American Heart Association.
\textit{Circulation}, \textbf{139}(10), e56--e528.
https://doi.org/10.1161/CIR.0000000000000659

\bibitem{fuster2014acute}
Fuster, V., \& Kovacic, J. C. (2014).
Acute coronary syndromes: Pathology, diagnosis, genetics, prevention, and treatment.
\textit{Circulation Research}, \textbf{114}(12), 1847--1851.
https://doi.org/10.1161/CIRCRESAHA.114.302562

\bibitem{mh20172017}
Yancy, C. W., Jessup, M., Bozkurt, B., Butler, J., Casey, D. E., Jr., Colvin, M. M., Drazner, M. H., Filippatos, G. S., Fonarow, G. C., Givertz, M. M., \textit{et al.} (2017).
2017 ACC/AHA/HFSA focused update of the 2013 ACCF/AHA guideline for the management of heart failure.
\textit{Circulation}, \textbf{136}, e137--e161.
https://doi.org/10.1161/CIR.0000000000000509

\bibitem{fowkes2013comparison}
Fowkes, F. G. R., Rudan, D., Rudan, I., Aboyans, V., Denenberg, J. O., McDermott, M. M., Norman, P. E., Sampson, U. K. A., Williams, L. J., Mensah, G. A., \textit{et al.} (2013).
Comparison of global estimates of prevalence and risk factors for peripheral artery disease in 2000 and 2010: A systematic review and analysis.
\textit{The Lancet}, \textbf{382}(9901), 1329--1340.
https://doi.org/10.1016/S0140-6736(13)61249-0

\bibitem{Roffi2016}
Roffi, M., Patrono, C., Collet, J.-P., \textit{et al.} (2016).
2015 ESC Guidelines for the management of acute coronary syndromes in patients presenting without persistent ST-segment elevation.
\textit{European Heart Journal}, \textbf{37}(3), 267--315.
https://doi.org/10.1093/eurheartj/ehv320

\bibitem{GerhardHerman2017}
Gerhard-Herman, M. D., Gornik, H. L., Barrett, C., \textit{et al.} (2017).
2016 AHA/ACC guideline on the management of patients with lower extremity peripheral artery disease.
\textit{Journal of the American College of Cardiology}, \textbf{69}(11), e71--e126.
https://doi.org/10.1016/j.jacc.2016.11.007

\bibitem{Go2013}
Go, A. S., Mozaffarian, D., Roger, V. L., \textit{et al.} (2013).
Heart disease and stroke statistics---2013 update: A report from the American Heart Association.
\textit{Circulation}, \textbf{127}(1), e6--e245.
https://doi.org/10.1161/CIR.0b013e31828124ad

\bibitem{Townsend2016}
Townsend, R. R., Wilkinson, I. B., Schiffrin, E. L., \textit{et al.} (2015).
Recommendations for improving and standardizing vascular research on arterial stiffness: A scientific statement from the American Heart Association.
\textit{Hypertension}, \textbf{66}(3), 698--722.
https://doi.org/10.1161/HYP.0000000000000033

\end{thebibliography}
\end{document}